\documentclass[
 twocolumn,
 superscriptaddress,
 showpacs,
 amsmath,
 amssymb,
 aps,
 pre
]{revtex4-1}

\usepackage[dvipdfmx]{graphicx}
\usepackage{dcolumn}
\usepackage{bm}

\usepackage{algcompatible}
\usepackage[size=small]{caption}
\DeclareCaptionFormat{algorithm}{\hrulefill\vskip-8pt\hrulefill\par#1#2#3\vskip-4pt\hrulefill}
\usepackage{etoolbox}
\AtEndEnvironment{algorithm}{\noindent\hrulefill\par\nobreak\vskip-8pt\hrulefill}
\usepackage{newfloat}
\DeclareFloatingEnvironment[
  fileext=loa,
  listname=List of Algorithms,
  name=ALGORITHM,
  placement=tbhp,
]{algorithm}
\usepackage{mathtools}
\usepackage{amsthm}

\usepackage{subcaption}
\usepackage{xcolor}
\usepackage{diagbox}
\usepackage{multirow}
\usepackage{xr}
\usepackage{braket}
\usepackage{amsmath}

\makeatletter
\newcommand{\vast}{\bBigg@{3.0}}
\newcommand{\Vast}{\bBigg@{3.5}}

\makeatother
\allowdisplaybreaks

\begin{document}

\preprint{APS/123-QED}

\title{Many-body perturbation theory and fluctuation relations \\ for interacting population dynamics}

\author{Hideyuki Miyahara}
\email{hideyuki\_miyahara@mist.i.u-tokyo.ac.jp}

\affiliation{%
Department of Mathematical Informatics,
Graduate School of Information Science and Technology,
The University of Tokyo,
Tokyo 113-8656, Japan
}%


\date{\today}


\begin{abstract}
Population dynamics deals with the collective phenomena of living organisms, and it has attracted much attention since it is expected to explain how not only living organisms but also human beings have been adapted to varying environments.
However, it is quite difficult to insist on a general statement on living organisms since mathematical models heavily depend on phenomena that we focus on.
Recently it was reported that the fluctuation relations on the fitness of living organisms held for a quite general problem setting.
But, interactions between organisms were not incorporated in the problem setting, though interaction plays critical roles in collective phenomena in physics and population dynamics.
In this paper, we propose interacting models for population dynamics and provide the perturbative theory of population dynamics.
Then, we derive the variational principle and fluctuation relations for interacting population dynamics.
\end{abstract}

%


\maketitle


\section{Introduction}

Population dynamics aims to describe the population growth of individuals that are able to multiply by themselves~\cite{Mayr01, Gavrilets01, Meszena01, Thieme01, Hofbauer01, Turchin01, Hartl01, Lande01, Donaldson01, Rivoire01, Rivoire02}.
Typical examples are organisms in living cells and human beings.
In the former case, they can increase their populations by cell division; for the latter case, they can multiply their numbers by giving birth.
To sustain life and to avoid extinction, the capability of multiplying via adapting a varying environment is essentially important for organisms and animals, respectively, and it critically distinguishes them from physical systems, such as condensed matter composed of electrons and spins.
In particular, the adaption of human beings to the varying environments is one of the biggest issues since Darwin's time~\cite{Mayr01, Gavrilets01, Meszena01}.

On the other hand, since the discoveries of Jarzynski's equality~\cite{Jarzynski01} and Crooks' relation~\cite{Crooks01}, the study of stochastic thermodynamics has attracted considerable attention~\cite{Seifert01}.
Recently, the relation between population dynamics and stochastic thermodynamics has been intensively studied, and several variants of the fluctuation relations (FRs) were discovered for the fitness of organisms in a general problem setting~\cite{Kobayashi03, Kobayashi04}.
However, there is a critical limitation in Refs.~\cite{Kobayashi03,Kobayashi04}.
The authors dealt with only one-body problems of organisms that can multiply into two following some processes; as a result, the population always grows or decays exponentially in the models studied in Refs.~\cite{Kobayashi03, Kobayashi04}.
On the other hand, models that do not show exponential growth, such as logistic growth models, ubiquitously appear in population dynamics~\cite{Tsoularis01}, and the FRs shown in Refs.~\cite{Kobayashi03, Kobayashi04} do not hold for them.

In this paper, we establish many-body perturbative theory~\cite{Abrikosov01, Fetter01} of interacting population dynamics and derive several FRs for interacting models in population dynamics.
To this end, we first propose a model that describes population dynamics with local interaction and derive a weakly interacting model by using the perturbation expansion.
Second, we formulate the perturbative theory of population dynamics and obtain the variational principle for interacting population dynamics.
Then, the variational principle with an optimal strategy leads to the consistency condition, which plays an essential role in deriving FRs for population dynamics.
Finally, we derive detailed FRs for interacting population dynamics.
We also obtain the Kawai-Parrondo-Broeck type FRs~\cite{Kawai01}.

This paper is organized as follows.
In Sec.~\ref{sec-models-00-01}, we introduce models with and without interaction and derive a model that is investigated in this paper by using the perturbation expansion.
At the end of this section, we discuss the validity of the model.
In Sec.~\ref{sec-variational-01}, we derive the variational principle for the model.
The variational principle gives another representation of the fitness and leads to a rich variety of mathematical relations.
In Sec.~\ref{sec-optimal-protocal-01}, we consider an optimal strategy and then derive a consistency condition for it.
We see that the variational principle leads to the consistency condition.
In Sec.~\ref{sec-FRs-01}, we derive several FRs.
In particular, we derive some detailed FRs and then the Kawai-Parrondo-Broeck type FRs.
We also explain that the consistency condition for the optimal strategy plays a central role in the FRs.
In Sec.~\ref{sec-second-law-01}, we derive an integral FRs and a second-law-like inequality for interacting population dynamics.
In Sec.~\ref{disc-01}, we discuss our findings and conclude this paper. In particular, we explain the meaning and limitations of them.
Furthermore, several proofs of the findings in Secs.~\ref{sec-variational-01}, \ref{sec-optimal-protocal-01}, and \ref{sec-FRs-01} are given in the Appendix.

%

\section{Models and its validity} \label{sec-models-00-01}

In this section, we introduce several models and explain the relations among them.
We begin with a noninteracting model for population dynamics and then introduce an interacting model.
Then, we consider the perturbative expansions of population growth, phenotype-switching, and hoping terms, and derive a model that is mainly investigated in this paper.
At the end of this section, we discuss the validity of this model with numerical simulation.

\subsection{Model without interaction} \label{sec-nonint-01}

In Refs.~\cite{Kobayashi03, Kobayashi04}, FRs for population dynamics were first established.
Here, we explain the model investigated in Refs.~\cite{Kobayashi03, Kobayashi04}, which does not involve interaction terms.

First, we define the basic setting and variables.
We consider a discrete-time model coupled with an environment and, since the biological processes on reproduction have periodicity in general.
We use $r_t$ and $f_t$ for the space coordinate and the phenotypic state, respectively, and put $x_t \coloneqq (r_t, f_t)$.
We also denote the state of the environment at $r_t$ by $y_t (r_t)$ and the history $\{y_i (r_i)\}_{i=1}^t$ by $Y_t (r_t)$.
Furthermore, let $N^{(t)} (x_t, Y_t (\cdot))$ be the number of organisms whose state is given by $x_t$ at time $t$, when the trajectory of the environment is given by $Y_t (\cdot)$.

The simplest model for population dynamics with phenotype switching has two terms: population growth, phenotype-switching, and hopping terms.
In Ref.~\cite{Kobayashi03} and the other literature, the following model was investigated:
\begin{align}
N^{(t)} (x_t, Y_t (\cdot))
&= D_0 (x_t, y_t (r_t)) \nonumber \\
& \quad \times \sum_{\{x_{t-1}\}} T_0 (x_t| x_{t-1}) \nonumber \\
& \quad \times N^{(t-1)} (x_{t-1}, Y_{t-1} (\cdot)). \label{model-01-01}
\end{align}
Note that Eq.~\eqref{model-01-01} focuses on the mean values of organisms by assuming that fluctuations around the mean values can be ignored.

The dynamics of population growth, which distinguishes population dynamics from other physical systems, such as electronic and magnetic systems, is expressed by $D_0 (\cdot, \cdot)$ (see Fig.~\ref{model-01-01}).
In general, multiplication tends to occur when resources are rich and the density of a species is low; however, this effect is ignored in Eq.~\eqref{model-01-01}.
This motivates us to consider the interaction effect on multiplication in this paper.

In the noninteracting case, $T_0 (x_t| x_{t-1})$ can be decomposed as
\begin{align}
  T_0 (x_t| x_{t-1}) &= T_0 (r_t| r_{t-1}) T_0 (f_t| f_{t-1}),
\end{align}
where $T_0 (r_t| r_{t-1})$ and $T_0 (f_t| f_{t-1})$ are noninteracting phenotype-switching and hopping terms, respectively.
An interaction effect on phenotype switching and hopping may also be important, but in this paper, we do not get into this problem.

\begin{figure}[t]
\centering
\includegraphics[scale=0.35]{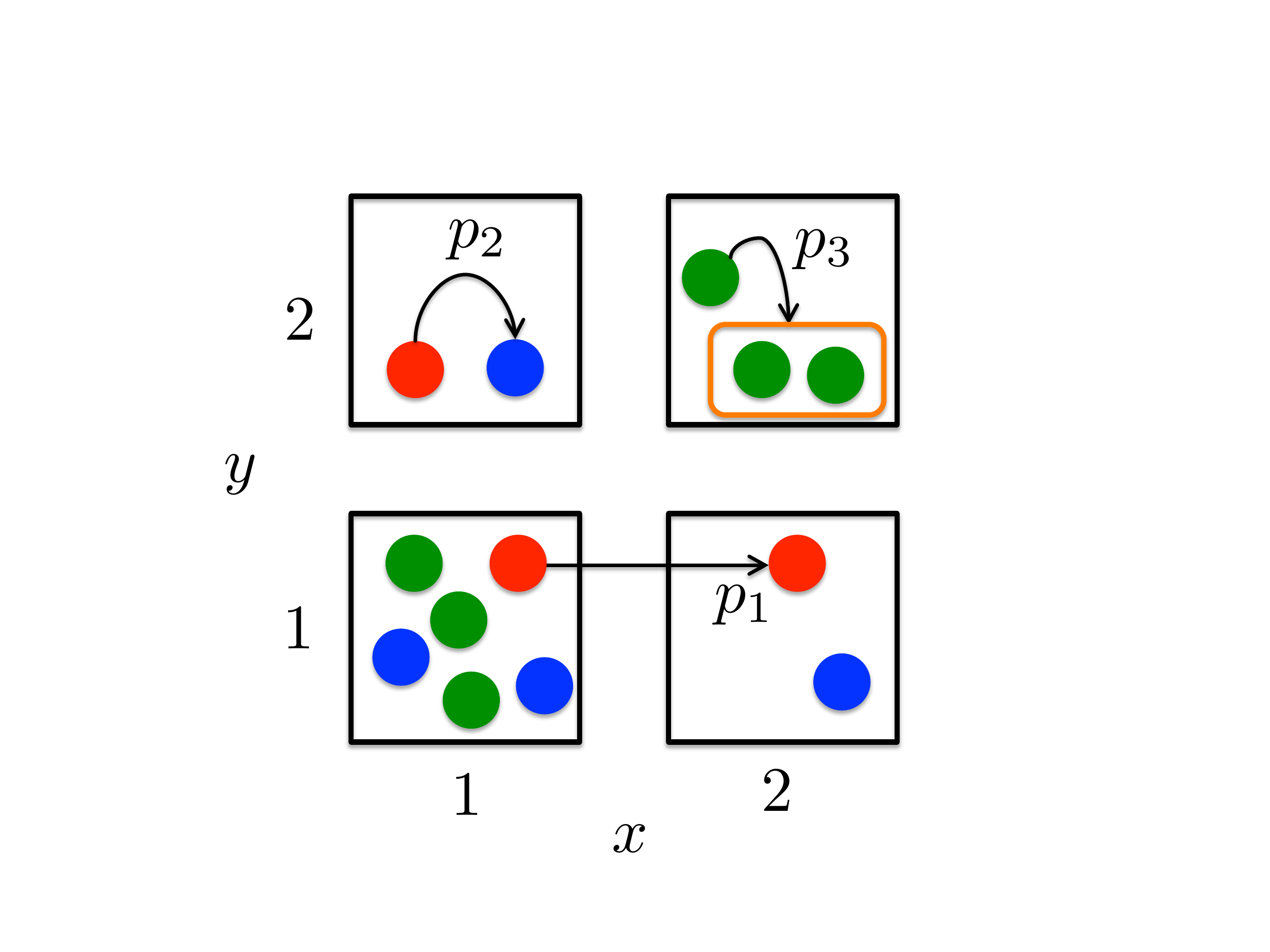}
\caption{Schematic of an example of the noninteracting model for population dynamics. This system has four sites and three phenotypes. This model has three processes: hopping $p_1$, phenotype switching (mutation) $p_2$, and population growth (multiplication) $p_3$. More specifically, $p_1$ represents hopping from $r_t = [1, 1]^\intercal$ to $r_{t+1} = [2, 1]^\intercal$, and $p_2$ represents phenotype switching from $f_t = \mathrm{``red"}$ to $f_{t+1} = \mathrm{``blue."}$ Furthermore, $p_3$ depicts multiplication of a cell whose phenotype is ``green". This dynamics is expressed by $D_0 (\cdot, \cdot)$.
}
\label{schematic-01-01}
\end{figure}

\subsection{Model with interaction} \label{sec-int-01}

We extend the noninteracting model~\eqref{model-01-01}, by introducing interaction effects on $D_0 (\cdot, \cdot)$ and $T_0 (\cdot| \cdot)$.
That is, we replace $D_0 (\cdot, \cdot)$ and $T_0 (\cdot| \cdot)$ by $D (\cdot, \cdot, \cdot)$ and $T (\cdot| \cdot, \cdot)$ at time $t$, which depend on $N^{(t-1)} (\cdot, \cdot)$, respectively, and then we obtain
\begin{align}
N^{(t)} (x_t, Y_t (\cdot))
&= D (x_t, N^{(t-1)} (\cdot, Y_{t-1} (\cdot)), y_t (r_t)) \nonumber \\
& \quad \times \sum_{\{x_{t-1}\}} T (x_t| x_{t-1}, N^{(t-1)} (\cdot, Y_{t-1} (\cdot))) \nonumber \\
& \quad \times N^{(t-1)} (x_{t-1}, Y_{t-1} (\cdot)). \label{model-01-02}
\end{align}
The point of Eq.~\eqref{model-01-02} is that the dependence of $D (\cdot, \cdot, \cdot)$ and $T (\cdot| \cdot, \cdot)$ at time $t$ on $N^{(t-1)} (\cdot, \cdot)$ can represent interaction effects, such as the excluded volume effect.

\subsection{Perturbation expansions of multiplication and phenotype-switching terms} \label{sec-perturbation-expansion-01}

In general, it is difficult to compute physical quantities on Eq.~\eqref{model-01-02}.
We then consider the perturbative expansions of $D (\cdot, \cdot, \cdot)$ and $T (\cdot| \cdot, \cdot)$:
\begin{align}
& D (x_t, N^{(t-1)} (\cdot, Y_{t-1} (\cdot)), y_t (r_t)) \nonumber \\
& \quad = D_0 (x_t, y_t (r_t)) \nonumber \\
& \qquad + \sum_{i=1}^\infty \sum_{\{x_{t-1}\}} D_i (x_t, x_{t-1}, y_t (r_t)) \nonumber \\
& \qquad \quad \times \Big( N^{(t-1)} (x_{t-1}, Y_{t-1} (\cdot)) \Big)^i, \label{taylor-01-01}
\end{align}
and
\begin{align}
& T (x_t| x_{t-1}, N^{(t-1)} (\cdot, Y_{t-1} (\cdot))) \nonumber \\
& \quad = T_0 (x_t| x_{t-1}) \nonumber \\
& \qquad + \sum_{i=1}^\infty \sum_{\{x_{t-1}'\}} T_i (x_t| x_{t-1}, x_{t-1}') \nonumber \\
& \qquad \quad \times \Big( N^{(t-1)} (x_{t-1}', Y_{t-1} (\cdot)) \Big)^i. \label{taylor-01-02}
\end{align}
Equations~\eqref{taylor-01-01} and \eqref{taylor-01-02} express nonlinear effects of population growth, phenotype switching, and hopping due to interaction that come from interaction.

\subsection{Model with weak interaction} \label{sec-weak-int-01}

So far, we have explained noninteracting and interacting models for population dynamics and the perturbation expansions.
Here, we introduce a model with weak interaction.
By considering the first-order expansion on $D (\cdot, \cdot, \cdot)$ and the zeroth-order expansion on $T (\cdot| \cdot, \cdot)$, we obtain
\begin{align}
N^{(t)} (x_t, Y_t (\cdot))
&= D (x_t, N^{(t-1)} (\cdot, Y_{t-1} (\cdot)), y_t (r_t)) \nonumber \\
& \quad \times \sum_{\{x_{t-1}\}} T_0 (x_t| x_{t-1}) \nonumber \\
& \quad \times N^{(t-1)} (x_{t-1}, Y_{t-1} (\cdot)), \label{model-01-03}
\end{align}
where $T_0 (\cdot| \cdot)$ is the transition matrix of phenotype switching and hopping, and the interaction term $D (\cdot, \cdot, \cdot)$ is written as
\begin{align}
& D (x_t, N^{(t-1)} (\cdot, Y_{t-1} (\cdot)), y_t (r_t)) \nonumber \\
& \quad = D_0 (x_t, y_t (r_t)) \nonumber \\
& \qquad + \sum_{\{x_{t-1}\}} D_1 (x_t, x_{t-1}, y_t (r_t)) N^{(t-1)} (x_{t-1}, Y_{t-1} (\cdot)). \label{growth-01}
\end{align}
Note that the first and second terms of the right-hand side of Eq.~\eqref{growth-01} represent one-body and interaction growth terms, respectively, and this model is almost the same with the model dealt in Refs.~\cite{Kobayashi03, Kobayashi04} if we set $D_1 (x_t, x_{t-1}, y_t (r_t)) = 0$.
Note that $N^{(0)} (\cdot, \cdot)$ is the population of the organisms at $t=0$.
Hereafter, we denote it by $N^{(0)} (\cdot)$ for simplicity since it does not depend on the state of the environment $Y_0 (\cdot)$.

\subsection{Validity of the model} \label{sec-model-validity-01}

We here discuss the validity of the model~\eqref{model-01-03}. 
The model~\eqref{model-01-03}, is based on the mean populations of each phenotype and higher-order cumulants of the populations, such as their variances, are assumed to be small enough.
Thus, the model is valid when each population is large~\cite{Rivoire01}.

Next, we turn our attention to the interaction in the model~\eqref{model-01-03}.
We incorporate the interaction effect only in the growth term; the reasons are as follows.
The first one is that the interaction effect on phenotype switching and hopping is similar to the interaction between spins in a spin model, such as the Ising model and the Potts model.
Thus, there are many works on it.
The second one is that when the number of organisms is larger, it is expected that organisms are less likely to multiply due to the exclusive volume effect and the exhaustion of resources.
And when an organism behaves like a catalyst, it promotes cell division.
This effect is essentially important to understand the collective phenomena of population dynamics.
The third one is that the interaction effect of the growth term can effectively describe the interaction effect on phenotype switching.

In Eq.~\eqref{model-01-03}, we have considered the time-delayed interaction represented by $D_1 (x_t, x_{t-1}, y_t (r_t))$.
The main reason is that cell division and other biological phenomena have periodicity in general, and it is natural to consider that there exists time delay.
On the other hand, the time-delayed interaction and a simultaneous interaction are perturbatively the same; thus, the results derived in this paper can be straightforwardly extended to a model with a simultaneous interaction.

\subsection{Numerical simulation} \label{sec-numerics-01}

Here, we demonstrate how the fitness of an interacting system behaves and compare its zeroth- and first-order approximations with it.

For simplicity, we fix the state of the environment and consider an interacting system that has one site and two phenotypes; so, we omit $r_t$ in this numerical simulation.
For phenotype switching, we set $T_0 (f_t = f_{t-1}| f_{t-1}) = 0.9$ and $T_0 (f_t \ne f_{t-1}| f_{t-1}) = 0.1$.
For population growth, we also put $D_0 (f_t = 1) = 1.10$, $D_0 (f_t = 2) = 1.02$, and $D_1 (f_t, f_t' = f_t) = - 0.010$.
We define $\bar{N}^{(t)}$ as the total population at time $t$.
The precise definition will be given in the next section.

\begin{figure}[t]
\centering
\includegraphics[scale=0.45]{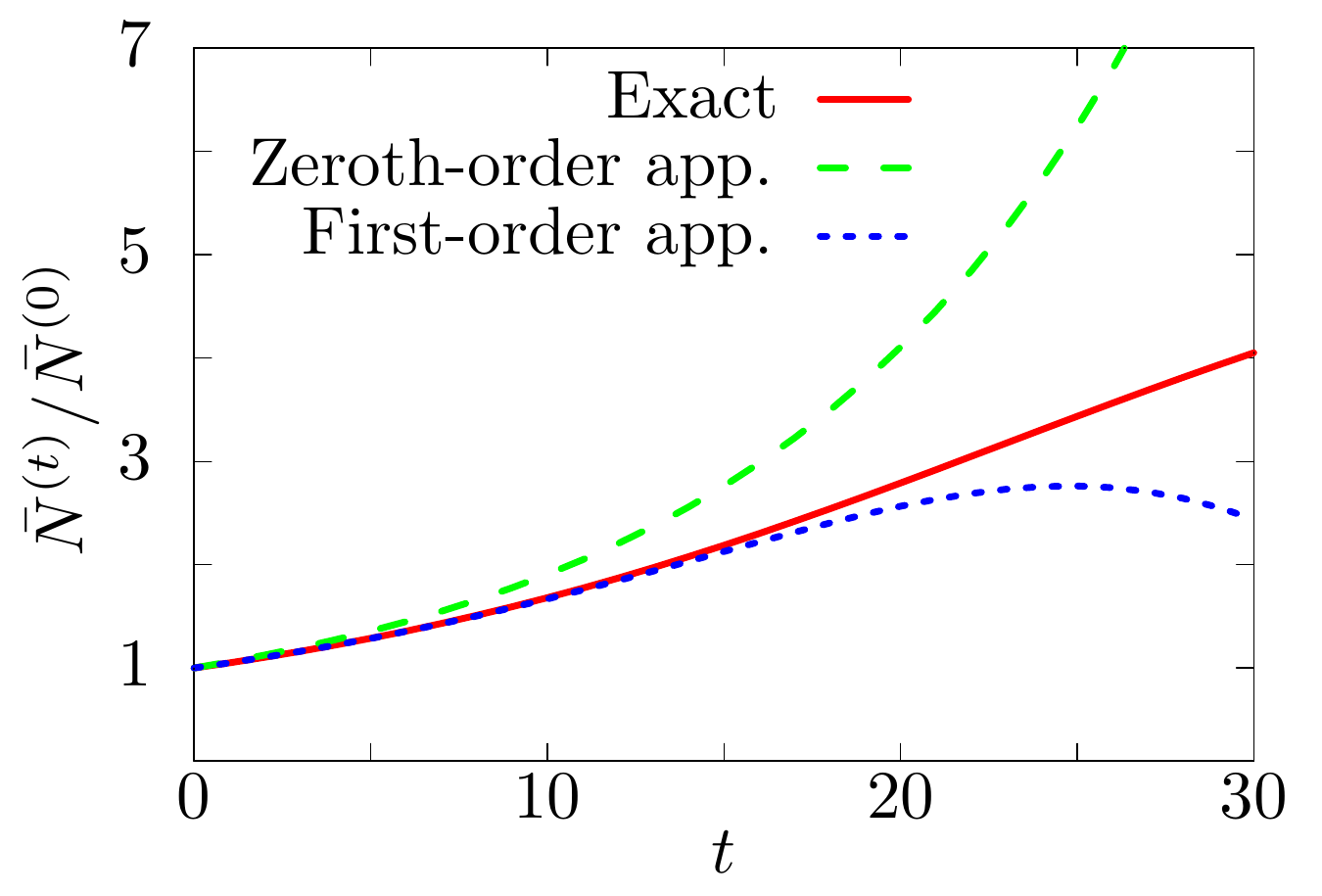}
\caption{Comparison of the exact computation (red), the zeroth-order perturbation approximation (green), and the first-order perturbation approximation (blue) of the fitness of an interacting system. We consider a system composed of one site and set $T_0 (f_t = f_{t-1}| f_{t-1}) = 0.9$, $T_0 (f_t \ne f_{t-1}| f_{t-1}) = 0.1$, $D_0 (f_t = 1) = 1.10$, $D_0 (f_t = 2) = 1.02$, and $D_1 (f_t, f_t' = f_t) = - 0.010$.}
\label{numerical-sim-01-01}
\end{figure}

In Fig.~\ref{numerical-sim-01-01}, we compare the exact result, the zeroth-order approximation, and the first-order approximation.
This figure shows that the exact result and the first-order approximation show good agreement with each other at the beginning while the zeroth-order approximation behaves in a different way even for the same time.
Due to the interaction effect, the exact result and the first-order approximation do not show an exponential growth; however, the zeroth-order approximation shows an exponential growth since it ignores interaction.

As Fig.~\ref{numerical-sim-01-01} also shows, the first-order approximation is valid at the beginning in this setup, because the population grows and higher-order terms become important as time elapses.
Thus, the first-order approximation is expected to be valid until higher-order terms dominate the system.

%


\section{Variational structure of interacting population dynamics} \label{sec-variational-01}

This section aims to derive the variational principle for interacting population dynamics, which provides another expression of the fitness of a population and makes it easy to derive a consistency condition for the optimal strategy.
To this end, this section begins with the definition of the log-fitness of a population and then states its path integral expression.
Finally, we derive the variational principle for interacting population dynamics.

\subsection{Log-fitness}

We here focus on $N^{(t)} (x_t, Y_t (\cdot))$ described by Eq.~\eqref{model-01-03}.
We then define the log fitness $\Phi_t^\mathrm{tot} (Y_t (\cdot))$, which describes how much the population grows in a given time, by
\begin{align}
\Phi_t^\mathrm{tot} (Y_t (\cdot)) &\coloneqq \ln \frac{\bar{N}^{(t)} (Y_t (\cdot))}{\bar{N}^{(0)}}. \label{log-fitness-02-01}
\end{align}
where $\bar{N}^{(t)} (Y_t (\cdot)) \coloneqq \sum_{\{x_t\}} N^{(t)} (x_t, Y_t (\cdot))$ for any $t \ge 1$ and $\bar{N}^{(0)} \coloneqq \sum_{\{x_0\}} N^{(0)} (x_0)$.
Here, $\sum_{\{x_t\}}$ represents the summation over all configurations of $x_t = (r_t, f_t)$.
Note that Eq.~\eqref{log-fitness-02-01} quantifies how much the total population grows logarithmically and does not depend on where organisms are and their phenotype.

\subsection{First-order perturbative expression}

We derive the path integral expression of the log-fitness~\eqref{log-fitness-02-01} within the first-order perturbation.
We first define the forward path probability $p_\mathrm{f} (X_t) \coloneqq \prod_{i=1}^t T_0 (x_i| x_{i-1}) p (x_0)$ with $p (x_0) \coloneqq N^{(0)} (x_0) / \bar{N}^{(0)}$ and $X_t \coloneqq \{ x_i \}_{i=0}^t$.

By using the first-order perturbation expansion, Eq.~\eqref{log-fitness-02-01} can be computed as
\begin{align}
\Phi_t^\mathrm{tot} (Y_t (\cdot)) &\approx \sum_{i=1}^t \ln \Big\langle \Big( D_0 (x_i, y_i (r_i)) \nonumber \\
& \quad + E_1^{(i, 0)} (x_i, y_i (r_i), Y_{i-1} (\cdot)) \bar{N}^{(0)} \Big) \Big\rangle_{p_\mathrm{f} (X_t)}, \label{first-order-perturbation-02-01}
\end{align}
where
\begin{align}
& E_1^{(i, 0)} (x_i, y_i (r_i), Y_{i-1} (\cdot)) \nonumber \\
& \quad \coloneqq \Bigg\langle D_1 (x_i, x_{i-1}, y_i (r_i)) \prod_{j=1}^{i-1} D_0 (x_j, y_j (r_j)) \Bigg\rangle_{p_\mathrm{f} (X_{i-1})},
\end{align}
for $i = 1, 2, \dots, t$.
We note that Eq.~\eqref{first-order-perturbation-02-01} has a similar structure with the Green's function in many-body systems~\cite{Abrikosov01, Fetter01}.
The details for the derivation of Eq.~\eqref{first-order-perturbation-02-01} are shown in Appendix~\ref{derivation-01-01}.
In the rest of this paper, we derive the variational principle and FRs by using Eq.~\eqref{first-order-perturbation-02-01}.
Hereafter, we use the equality when two quantities are perturbatively equal.

\subsection{Variational principle}

Then, we derive the variational principle on the log-fitness~\eqref{log-fitness-02-01}.
It plays an important role in this paper since it leads to the FRs shown later.

By applying Jensen's inequality to Eq.~\eqref{first-order-perturbation-02-01}, we obtain the inequality on Eq.~\eqref{log-fitness-02-01}:
\begin{align}
\Phi_t^\mathrm{tot} (Y_t (\cdot)) &\ge \sum_{i=1}^{t} \Big\langle \ln \Big( D_0 (x_i, y_i (r_i)) \nonumber \\
& \quad \qquad + \tilde{E}_1 (X_i, Y_i (\cdot)) \bar{N}^{(0)} \Big) \Big\rangle_{q (X_i)} \nonumber \\
& \qquad - \sum_{i=1}^{t} \mathrm{KL} \Big( q(X_i) \Big\| p_\mathrm{f} (X_i) \Big), \label{lower-bound-fitness-02-01}
\end{align}
where
\begin{align}
& \tilde{E}_1 (X_i, Y_i (\cdot)) \nonumber \\
& \quad \coloneqq D_1 (x_i, x_{i-1}, y_i (r_i)) \Bigg[ \prod_{j=1}^{i-1} D_0 (x_j, y_j (r_j)) \Bigg], \label{interaction-term-path-02-02}
\end{align}
for any set of path measures $\{ q(X_i) \}_{i=0}^t$.
See Appendix~\ref{derivation-01-02} for details.

Next, we consider the equality condition of Eq.~\eqref{lower-bound-fitness-02-01}.
We here define the backward path probabilities by
\begin{align}
p_\mathrm{b} (X_j| Y_j (\cdot)) &\coloneqq \Big( D_0 (x_j, y_j (r_j)) + \tilde{E}_1 (X_j, Y_j (\cdot)) \bar{N}^{(0)} \Big) \nonumber \\
& \quad \times e^{- \Phi_j (Y_j (\cdot))} p_\mathrm{f} (X_j), \label{backward-path-probability-01}
\end{align}
where
\begin{align}
& \Phi_j (Y_j (\cdot)) \nonumber \\
& \quad \coloneqq \ln \Braket{ \Big( D_0 (x_j, y_j (r_j)) + \tilde{E}_1 (X_j, Y_j (\cdot)) \bar{N}^{(0)} \Big) }_{p_\mathrm{f} (X_j)}, \label{partial-log-fitness-02-01-01}
\end{align}
for $j = 1, 2, \dots, t$.
Note that $\Phi_j (Y_j (\cdot))$ in Eq.~\eqref{partial-log-fitness-02-01-01} and $\Phi_t^\mathrm{tot} (Y_t (\cdot))$ in Eq.~\eqref{log-fitness-02-01} satisfy
\begin{align}
\Phi_t^\mathrm{tot} (Y_t (\cdot)) &= \sum_{j=1}^t \Phi_j (Y_j (\cdot)).
\end{align}
Then, we have
\begin{align}
\Phi_t^\mathrm{tot} (Y_t (\cdot)) &= \sum_{i=1}^t \Big\langle \ln \Big( D_0 (x_i) \nonumber \\
& \qquad + \tilde{E}_1 (X_i, Y_i (\cdot)) \bar{N}^{(0)} \Big) \Big\rangle_{p_\mathrm{b} (X_i| Y_i (\cdot))} \nonumber \\
& \qquad - \sum_{i=1}^t \mathrm{KL} \Big( p_\mathrm{b} (X_i| Y_i (\cdot)) \Big\| p_\mathrm{f} (X_i) \Big). \label{variational-representation-02-01}
\end{align}
We note that Eq.~\eqref{variational-representation-02-01} represents the relation between the log fitness and the forward and backward path probabilities.
See Appendix~\ref{derivation-01-03} for details.

As a result, we have the variational representation of the log fitness given by
\begin{align}
& \Phi_t^\mathrm{tot} (Y_t (\cdot)) \nonumber \\
& \quad = \max_{\{ q (X_i) \}_{i=i}^t} \nonumber \\
& \qquad \Bigg[ \sum_{i=1}^t \Big\langle \ln \Big( D_0 (x_i) + \tilde{E}_1 (X_i, Y_i (\cdot)) \bar{N}^{(0)} \Big) \Big\rangle_{q (X_i)} \nonumber \\
& \qquad - \sum_{i=1}^t \mathrm{KL} \Big( q(X_i) \Big\| p_\mathrm{f} (X_i) \Big) \Bigg]. \label{max-variational-representation-02-01}
\end{align}

%

\section{Optimal protocol} \label{sec-optimal-protocal-01}

In this section, we consider the optimal protocol of phenotype switching.
By using the nature of optimality, we derive a consistency condition of the optimal protocol on the path probabilities on the forward and backward processes.
The consistency condition plays an essential role in FRs in the next section.

We first consider the expectation of the log-fitness with respect to the states of the environment and then derive the consider condition by utilizing the nature of optimality.
In addition, we find a variational principle for the optimal strategy.

\subsection{Derivation of the fitness}

We here derive the deviation of the fitness to consider properties of the optimal protocol and stochastic thermodynamic structure~\cite{Jarzynski01, Jarzynski02, Jarzynski03, Crooks01, Seifert01, Seifert02, Seifert03, Kurchan01, Miyahara07} on population dynamics.

Let us write the path probability of the environment by $p_\mathrm{e} (Y_t (\cdot))$.
The expectation of $\Phi_t^\mathrm{tot} (Y_t (\cdot))$ in Eq.~\eqref{variational-representation-02-01} with respect to $p_\mathrm{e} (Y_t (\cdot))$ is expressed as
\begin{align}
& \Braket{ \Phi_t^\mathrm{tot} (Y_t (\cdot)) }_{p_\mathrm{e} (Y_t (\cdot))} \nonumber \\
& \quad = \sum_{i=1}^t \Big\langle \ln \Big( D_0 (x_i) \nonumber \\
& \qquad + \tilde{E}_1 (X_i, Y_i (\cdot)) \bar{N}^{(0)} \Big) \Big\rangle_{p_\mathrm{b} (X_i, Y_i (\cdot))} \nonumber \\
& \qquad - \sum_{i=1}^t \Big[ I_\mathrm{b}^{X_i, Y_i} + \mathrm{KL} \Big( p_\mathrm{b} (X_i) \Big\| p_\mathrm{f} (X_i) \Big) \Big], \label{expectation-Phi-02-02}
\end{align}
where
\begin{align}
I_\mathrm{b}^{X_i, Y_i} &\coloneqq \Bigg\langle \ln \frac{p_\mathrm{b} (X_i, Y_i (\cdot))}{p_\mathrm{b} (X_i) p_\mathrm{e} (Y_i (\cdot))} \Bigg\rangle_{p_\mathrm{b} (X_i, Y_i (\cdot))}.
\end{align}
Furthermore, we have defined
\begin{align}
p_\mathrm{b} (X_i, Y_i (\cdot)) &\coloneqq p_\mathrm{b} (X_i| Y_i (\cdot)) p_\mathrm{e} (Y_i (\cdot)),
\end{align}
and
\begin{align}
p_\mathrm{b} (X_i) &\coloneqq \Braket{ p_\mathrm{b} (X_i, Y_i (\cdot)) }_{p_\mathrm{e} (Y_i (\cdot))},
\end{align}
for $i = 1, 2, \dots, t$.
See Appendix~\ref{derivation-02-01} for details.

Then, we consider the deviation of the fitness from the optimal one.
We then define
\begin{align}
& \delta \Phi_t^\mathrm{tot} (Y_t (\cdot)) \nonumber \\
& \quad \coloneqq \sum_{i=1}^{t} \ln \Big\langle \Big( D_0 (x_i, y_i (r_i)) \nonumber \\
& \quad \qquad + \tilde{E}_1 (X_i, Y_i (\cdot)) \bar{N}^{(0)} \Big) \Big\rangle_{p_\mathrm{f} (X_i) + \delta p_\mathrm{f} (X_i) } \nonumber \\
& \qquad - \sum_{i=1}^{t} \ln \Big\langle \Big( D_0 (x_i, y_i (r_i)) \nonumber \\
& \quad \qquad + \tilde{E}_1 (X_i, Y_i (\cdot)) \bar{N}^{(0)} \Big) \Big\rangle_{p_\mathrm{f} (X_i)}. \label{def-dev-fitness-01}
\end{align}
Due to the fact that $\Phi_t^\mathrm{tot} (Y_t (\cdot))$ satisfies the maximization formula, Eq.~\eqref{max-variational-representation-02-01}, we have
\begin{align}
\delta \Phi_t^\mathrm{tot} (Y_t (\cdot)) &= \sum_{i=1}^t \Bigg\langle \frac{p_\mathrm{b} (X_i| Y_i (\cdot))}{p_\mathrm{f} (X_i)} \Bigg\rangle_{\delta p_\mathrm{f} (X_i)}. \label{delta-Phi-01}
\end{align}
Then, by taking the expectation of the left-hand side of Eq.~\eqref{delta-Phi-01} with respect to $p_\mathrm{e}$ we have
\begin{align}
\Braket{\delta \Phi_t^\mathrm{tot} (Y_t (\cdot))}_{p_\mathrm{e} (Y_t (\cdot))} &= \sum_{i=1}^t \Bigg\langle \frac{p_\mathrm{b} (X_i)}{p_\mathrm{f} (X_i)} \Bigg\rangle_{\delta p_\mathrm{f} (X_i)}, \label{delta-Phi-02}
\end{align}
where
\begin{align}
p_\mathrm{b} (X_i) &\coloneqq \sum_{\{Y_i (\cdot)\}} p_\mathrm{b} (X_i| Y_i (\cdot)) p_\mathrm{e} (Y_i (\cdot)),
\end{align}
for $i = 1, 2, \dots, t$.
See Appendix~\ref{derivation-02-02} for details.

\subsection{Optimal protocol.}

Next, we discuss the optimal strategy and the corresponding fitness $\hat{\Phi}_t^\mathrm{tot}$.
To this end, by letting $\hat{p}_\mathrm{f} (X_t)$ be the optimal forward path probability, we define the optimal backward path probability $\hat{p}_\mathrm{b} (X_t)$ by
\begin{align}
\hat{p}_\mathrm{b} (X_j| Y_j (\cdot)) &\coloneqq \Big( D_0 (x_j, y_j (r_j)) + \tilde{E}_1 (X_j, Y_j (\cdot)) \bar{N}^{(0)} \Big) \nonumber \\
& \quad \times e^{- \hat{\Phi}_j (Y_j (\cdot))} \hat{p}_\mathrm{f} (X_j),
\end{align}
where
\begin{align}
& \hat{\Phi}_j (Y_j (\cdot)) \nonumber \\
& \quad \coloneqq \ln \Braket{ \Big( D_0 (x_j, y_j (r_j)) + \tilde{E}_1 (X_j, Y_j (\cdot)) \bar{N}^{(0)} \Big) }_{\hat{p}_\mathrm{f} (X_j)}, \label{log-fitness-hat-02-03-01}
\end{align}
for $j = 1, 2, \dots, t$.
Like Eq.~\eqref{first-order-perturbation-02-01}, we also define
\begin{align}
\hat{\Phi}_t^\mathrm{tot} (Y_t (\cdot)) &\coloneqq \sum_{i=1}^{t} \ln \Big\langle \Big( D_0 (x_i, y_i (r_i)) \nonumber \\
& \quad + \tilde{E}_1 (X_i, Y_i (\cdot)) \bar{N}^{(0)} \Big) \Big\rangle_{\hat{p}_\mathrm{f} (X_i)}. \label{log-fitness-hat-02-03-02}
\end{align}
Note that $\hat{\Phi}_j (Y_j (\cdot))$ in Eq.~\eqref{log-fitness-hat-02-03-01} and $\hat{\Phi}_t^\mathrm{tot} (Y_t (\cdot))$ in Eq.~\eqref{log-fitness-hat-02-03-02} satisfy
\begin{align}
\hat{\Phi}_t^\mathrm{tot} (Y_t (\cdot)) &= \sum_{j=1}^t \hat{\Phi}_j (Y_j (\cdot)). \label{sum-rule-02-02-02}
\end{align}

The optimality condition is expressed as
\begin{align}
\Braket{\delta \hat{\Phi}_t^\mathrm{tot} (Y_t (\cdot))}_{p_\mathrm{e} (Y_t (\cdot))} &= 0. \label{opt-cond-02-01}
\end{align}
Equation~\eqref{opt-cond-02-01} is satisfied via
\begin{align}
\hat{p}_\mathrm{b} (X_i) &= \hat{p}_\mathrm{f} (X_i), \label{opt-cond-02-02}
\end{align}
for $i = 1, 2, \dots, t$.
In this case, we have
\begin{align}
& \Braket{ \hat{\Phi}_t^\mathrm{tot} (Y_t (\cdot)) }_{p_\mathrm{e} (Y_t (\cdot))} \nonumber \\
& \quad = \sum_{i=1}^{t} \Big\langle \ln \Big( D_0 (x_i, y_i (r_i)) \nonumber \\
& \qquad + \tilde{E}_1 (X_i, Y_i (\cdot)) \bar{N}^{(0)} \Big) \Big\rangle_{\hat{p}_\mathrm{b} (X_i, Y_i (\cdot))} - \sum_{i=1}^t \hat{I}_\mathrm{b}^{X_i, Y_i},
\end{align}
where
\begin{align}
\hat{I}_\mathrm{b}^{X_i, Y_i} &\coloneqq \Bigg\langle \ln \frac{\hat{p}_\mathrm{b} (X_i, Y_i (\cdot))}{\hat{p}_\mathrm{b} (X_i) p_\mathrm{e} (Y_i (\cdot))} \Bigg\rangle_{\hat{p}_\mathrm{b} (X_i, Y_i (\cdot))}.
\end{align}

We can also express $\Braket{\hat{\Phi}_t^\mathrm{tot} (Y_t (\cdot))}_{p_\mathrm{e} (Y_t (\cdot))}$ as
\begin{align}
& \Braket{ \hat{\Phi}_t^\mathrm{tot} (Y_t (\cdot)) }_{p_\mathrm{e} (Y_t (\cdot))} \nonumber \\
& \quad = \max_{\{ q (X_i| Y_i (\cdot)) \}_{i=1}^t} \nonumber \\
& \qquad \Bigg[ \sum_{i=1}^{t} \Big\langle \ln \Big( D_0 (x_i, y_i (r_i)) \nonumber \\
& \quad \qquad + \tilde{E}_1 (X_i, Y_i (\cdot)) \bar{N}^{(0)} \Big) \Big\rangle_{q (X_i| Y_i (\cdot)) p_\mathrm{e} (Y_i (\cdot))} \nonumber \\
& \quad \qquad - \sum_{i=1}^{t} I^{X_i, Y_i} \Bigg]. \label{maximization-principle-02-10-02}
\end{align}
where
\begin{align}
I^{X_i, Y_i} &\coloneqq \Bigg\langle \ln \frac{q (X_i, Y_i (\cdot))}{q (X_i) p_\mathrm{e} (Y_i (\cdot))} \Bigg\rangle_{q (X_i, Y_i (\cdot))},
\end{align}
and
\begin{align}
q (X_i, Y_i (\cdot)) &\coloneqq q (X_i| Y_i (\cdot)) p_\mathrm{e} (Y_i (\cdot)), \\
q (X_i) &\coloneqq \Braket{q (X_i, Y_i (\cdot))}_{p_\mathrm{e} (Y_i (\cdot))}.
\end{align}
We have shown the variational principle for $\Braket{ \hat{\Phi}_t^\mathrm{tot} (Y_t (\cdot)) }_{p_\mathrm{e} (Y_t (\cdot))}$.

%

\section{Fluctuation relations} \label{sec-FRs-01}

This section is the main part of this paper, in which we derive several FRs for interacting population dynamics.
At first, we derive detailed FRs.
These FRs resemble conventional FRs in stochastic thermodynamics~\cite{Seifert01}.
Then, we derive Kawai-Parrondo-Broeck type FRs~\cite{Kawai01}.

\subsection{Detailed FRs}

We define the deviation of the log fitness $\hat{\Phi}_t^\mathrm{tot} (Y_t (\cdot))$ as
\begin{align}
\Delta \hat{\Phi}_t^\mathrm{tot} (Y_t (\cdot)) &\coloneqq \hat{\Phi}_t^\mathrm{tot} (Y_t (\cdot)) - \Phi_t^\mathrm{tot} (Y_t (\cdot)).
\end{align}
We then have the following FR given by
\begin{align}
e^{- \Delta \hat{\Phi}_t^\mathrm{tot} (Y_t (\cdot))} &= \prod_{j=1}^t \frac{\hat{p}_\mathrm{b} (X_j| Y_j (\cdot))}{\hat{p}_\mathrm{f} (X_j)} \frac{p_\mathrm{f} (X_j)}{p_\mathrm{b} (X_j| Y_j (\cdot))}. \label{FR-main-01}
\end{align}
Furthermore, Eq.~\eqref{FR-main-01} can be rewritten as
\begin{align}
e^{- \Delta \hat{\Phi}_t^\mathrm{tot} (Y_t (\cdot))} &= \prod_{j=1}^t \frac{\hat{p}_\mathrm{b} (Y_j (\cdot)| X_j) p_\mathrm{f} (X_j)}{p_\mathrm{b} (X_j, Y_j (\cdot))}, \label{fluctuation-relation-main-02-01}
\end{align}
and
\begin{align}
e^{- \Delta \hat{\Phi}_t^\mathrm{tot} (Y_t (\cdot))} &= \prod_{j=1}^t \frac{\Big\langle \hat{p}_\mathrm{b} (Y_j (\cdot)| X_j) \Big\rangle_{p_\mathrm{f} (X_j)}}{p_\mathrm{e} (Y_j (\cdot))}. \label{fluctuation-relation-main-02-02}
\end{align}
Note that we have used Eq.~\eqref{opt-cond-02-02} to derive Eqs.~\eqref{fluctuation-relation-main-02-01} and \eqref{fluctuation-relation-main-02-02}.
See Appendix~\ref{derivation-03-01}.

\subsection{Kawai-Parrondo-Broeck type FRs}

We also have the Kawai-Parrondo-Broeck type FRs~\cite{Kawai01} represented by
\begin{align}
& \overline{\Braket{\Delta \hat{\Phi}_t^\mathrm{tot} (Y_t (\cdot))}} \nonumber \\
& \quad = \mathrm{KL} \Big( p_\mathrm{e} (Y_j (\cdot)) \Big\| \Braket{\hat{p}_\mathrm{b} (Y_j (\cdot)| X_j)}_{p_\mathrm{F} (X_j)} \Big),
\end{align}
\begin{align}
& \overline{\Braket{\Delta \hat{\Phi}_t^\mathrm{tot} (Y_t (\cdot))}} \nonumber \\
& \quad = \sum_{j=1}^t \mathrm{KL} \Big( \hat{p}_\mathrm{f} (X_j) \Big\| p_\mathrm{f} (X_j) \Big) \nonumber \\
& \qquad - \sum_{j=1}^t \Big\langle \mathrm{KL} \Big( \hat{p}_\mathrm{b} (X_j| Y_j (\cdot)) \Big\| p_\mathrm{b} (X_j| Y_j (\cdot)) \Big) \Big\rangle_{p_\mathrm{e} (Y_j (\cdot))}, \label{fluctuation-relation-kawai-02-01}
\end{align}
and
\begin{align}
& \overline{\Braket{\Delta \hat{\Phi}_t^\mathrm{tot} (Y_t (\cdot))}} \nonumber \\
& \quad = \sum_{j=1}^t \mathrm{KL} \Big( \hat{p}_\mathrm{f} (X_j) \Big\| p_\mathrm{f} (X_j) \Big) \nonumber \\
& \qquad - \sum_{j=1}^t \mathrm{KL} \Big( \hat{p}_\mathrm{b} (X_j, Y_j (\cdot)) \Big\| p_\mathrm{b} (X_j, Y_j (\cdot)) \Big), \label{fluctuation-relation-kawai-02-02}
\end{align}
where
\begin{align}
\overline{\Braket{\Delta \hat{\Phi}_t^\mathrm{tot} (Y_t (\cdot))}} 
&\coloneqq \sum_{j=1}^t \Braket{\Delta \hat{\Phi}_j (Y_j (\cdot))}_{p_\mathrm{e} (Y_j (\cdot))}. \label{total-log-fitness-expectation-01-01}
\end{align}
Through Eqs~\eqref{fluctuation-relation-main-02-01}, \eqref{fluctuation-relation-main-02-02}, \eqref{fluctuation-relation-kawai-02-01}, and \eqref{fluctuation-relation-kawai-02-02}, we have proved a variety of FRs for interacting population dynamics.
See Appendix~\ref{derivation-03-03} for details.

%

\section{Integral FRs and second-law-like inequalities} \label{sec-second-law-01}

Finally, we mention that, from Eq.~\eqref{fluctuation-relation-main-02-02}, we can easily derive integral fluctuation relations and second-law-like inequalities that characterize the efficiencies of the optimal strategy and another.

We first define
\begin{align}
\overline{\Braket{e^{- \Delta \hat{\Phi}_t^\mathrm{tot} (Y_t (\cdot))}}} &\coloneqq \prod_{j=1}^t \Braket{e^{- \Delta \hat{\Phi}_j (Y_j (\cdot))}}_{p_\mathrm{e} (Y_j (\cdot))}. \label{total-log-fitness-expectation-01-02}
\end{align}
Then, we have the Jarzynski-type equality
\begin{align}
\overline{\Braket{e^{- \Delta \hat{\Phi}_t^\mathrm{tot} (Y_t (\cdot))}}} &= 1. \label{Jarzynski-type-eq-01-01}
\end{align}
By applying Jensen's inequality to Eq.~\eqref{Jarzynski-type-eq-01-01}, we obtain the following second-law-like inequality for  Eq.~\eqref{total-log-fitness-expectation-01-01}:
\begin{align}
\overline{\Braket{\Delta \hat{\Phi}_t^\mathrm{tot} (Y_t (\cdot))}} &\ge 0.
\end{align}
In the noninteracting limit, these relations recover relations shown in Ref.~\cite{Kobayashi03}.


%
%
%

%

\section{Discussion and conclusion} \label{disc-01}

In this paper, we have derived various types of FRs on interacting population dynamics.
In the previous works~\cite{Kobayashi03, Kobayashi04}, the interaction effect was ignored, but it is widely believed that interaction plays a critical role in statistical mechanics.
Thus, the most important point of this paper is that we have dealt with an interacting model for population dynamics, which is expected to cover a wide range of models in population dynamics.
For instance, the SIR model is one of the most famous models with nonlinear terms~\cite{Kermack01}.
The origin of the nonlinear terms is the interactions among susceptible, infected, and recovered individuals.
%
Furthermore, without interaction, a model of population dynamics always shows exponential growth.
However, in most cases, it is not realistic; otherwise, the system would be governed by the species and the model would be broken down.

In Ref.~\cite{Kobayashi03}, some properties of FRs are discussed.
One of the most important properties is that suboptimal strategies may outperform the optimal strategy due to fluctuations of the environment.
Our FRs also insist that the above statement holds even if an interaction exists.
In the noninteracting limit, the FRs found in this paper are identical with those in Ref.~\cite{Kobayashi03}; so, our findings are viewed as a direct extension of FRs in Ref.~\cite{Kobayashi03}.

Finally, we discuss issues that we have not tackled in this paper.
First, we have not discussed the capability of each organism to sense the state of the environment.
However, by incorporating it in the phenotype-switching and hopping rate $T_0 (\cdot| \cdot)$, we can directly extend the framework and the FRs in this paper by following Refs.~\cite{Rivoire01, Kobayashi03, Kobayashi04}.
Second, we have considered only the first-order correction.
But, our framework can be generalized straightforwardly to include higher-order perturbation corrections.
Another issue in interacting population dynamics is the interaction effect on the phenotype-switching and hopping rate $T_0 (\cdot| \cdot)$.
This issue may lead to another modification; so, this is one of our future work.

%


\section*{Acknowledgements} \label{ack-00}

H.M. thanks T. J. Kobayashi and K. Aihara for fruitful discussions.
H.M. is supported by JSPS KAKENHI Grant No. JP18J12175.

%
%
%

%

\appendix

\begin{widetext}
\section{Derivations of the perturbative expression of the log-fitness and the variational principle}

By employing perturbation theory~\cite{Abrikosov01, Fetter01}, we provide the detailed derivations of Eqs.~\eqref{first-order-perturbation-02-01}, \eqref{lower-bound-fitness-02-01}, and \eqref{variational-representation-02-01} in this appendix.

%

\subsection{Derivation of Eq.~\eqref{first-order-perturbation-02-01}} \label{derivation-01-01}

Here, we provide the detailed derivation of Eq.~\eqref{first-order-perturbation-02-01}.
First, we compute the exact relation between $N^{(t)} (x_t, Y_t (\cdot))$ and $N^{(t-2)} (x_{t-2}, Y_{t-2} (\cdot))$ by recursively using Eq.~\eqref{model-01-03}.
Then, we derive the first-order perturbative relation between $N^{(t)} (x_t, Y_t (\cdot))$ and $N^{(t-2)} (x_{t-2}, Y_{t-2} (\cdot))$ and that between $N^{(t)} (x_t, Y_t (\cdot))$ and $N^{(t-3)} (x_{t-3}, Y_{t-3} (\cdot))$ by ignoring higher-order terms with respect to interaction $D_1$.
As a result, we straightforwardly get the first-order perturbative relation between $N^{(t)} (x_t, Y_t (\cdot))$ and $N^{(0)} (x_0, Y_0 (\cdot))$.

\subsubsection{Exact relation between $N^{(t)} (x_t, Y_t (\cdot))$ and $N^{(t-2)} (x_{t-2}, Y_{t-2} (\cdot))$}


To obtain the path integral formulation of the fitness of the population dynamics, Eq.~\eqref{first-order-perturbation-02-01}, we need the relation between $N^{(t)} (\cdot, \cdot)$ and $N^{(0)} (\cdot, \cdot)$.
For the first step, by recursively inserting Eq.~\eqref{model-01-03}, we have the exact relation between $N^{(t)} (\cdot, \cdot)$ and $N^{(t-2)} (\cdot, \cdot)$ given by
\begin{align}
N^{(t)} (x_t, Y_t (\cdot)) &= D_0 (x_t, y_t (r_t)) \sum_{\{x_{t-1}\}} T_0 (x_t| x_{t-1}) D_0 (x_{t-1}, y_{t-1} (r_{t-1})) \sum_{\{x_{t-2}\}} T_0 (x_{t-1}| x_{t-2}) N^{(t-2)} (x_{t-2}, Y_{t-2} (\cdot)) \nonumber \\
& \quad + D_0 (x_t, y_t (r_t)) \sum_{\{x_{t-1}\}} T_0 (x_t| x_{t-1}) E_1^{(t-1, t-2)} (x_{t-1}, y_{t-1} (r_{t-1}), Y_{t-2} (\cdot)) \bar{N}^{(t-2)} (Y_{t-2} (\cdot)) \nonumber \\
& \qquad \times \sum_{\{x_{t-2}\}} T_0 (x_{t-1}| x_{t-2}) N^{(t-2)} (x_{t-2}, Y_{t-2} (\cdot)) \nonumber \\
& \quad + E_1^{(t, t-2)} (x_t, y_t (r_t), Y_{t-1} (\cdot)) \bar{N}^{(t-2)} (Y_{t-2} (\cdot)) \sum_{\{x_{t-1}\}} T_0 (x_t| x_{t-1}) D_0 (x_{t-1}, y_{t-1} (r_{t-1})) \nonumber \\
& \qquad \times \sum_{\{x_{t-2}\}} T_0 (x_{t-1}| x_{t-2}) N^{(t-2)} (x_{t-2}, Y_{t-2} (\cdot)) \nonumber \\
& \quad + E_2^{(t, t-1, t-2)} (x_t, y_t (r_t), Y_{t-1} (\cdot)) \Big( \bar{N}^{(t-2)} (Y_{t-2} (\cdot)) \Big)^2 \sum_{\{x_{t-1}\}} T_0 (x_t| x_{t-1}) D_0 (x_{t-1}, y_{t-1} (r_{t-1})) \nonumber \\
& \qquad \times \sum_{\{x_{t-2}\}} T_0 (x_{t-1}| x_{t-2}) N^{(t-2)} (x_{t-2}, Y_{t-2} (\cdot)) \nonumber \\
& \quad + E_1^{(t, t-2)} (x_t, y_t (r_t), Y_{t-1} (\cdot)) \bar{N}^{(t-2)} (Y_{t-2} (\cdot)) \nonumber \\
& \qquad \times \sum_{\{x_{t-1}\}} T_0 (x_t| x_{t-1}) E_1^{(t-1, t-2)} (x_{t-1}, y_{t-1} (r_{t-1}), Y_{t-2} (\cdot)) \bar{N}^{(t-2)} (Y_{t-2} (\cdot)) \nonumber \\
& \qquad \times \sum_{\{x_{t-2}\}} T_0 (x_{t-1}| x_{t-2}) N^{(t-2)} (x_{t-2}, Y_{t-2} (\cdot)) \nonumber \\
& \quad + E_2^{(t, t-1, t-2)} (x_t, y_t (r_t), Y_{t-1} (\cdot)) \Big( \bar{N}^{(t-2)} (Y_{t-2} (\cdot)) \Big)^2 \nonumber \\
& \qquad \times \sum_{\{x_{t-1}\}} T_0 (x_t| x_{t-1}) E_1^{(t-1, t-2)} (x_{t-1}, y_{t-1} (r_{t-1}), Y_{t-2} (\cdot)) \bar{N}^{(t-2)} (Y_{t-2} (\cdot)) \nonumber \\
& \qquad \times \sum_{\{x_{t-2}\}} T_0 (x_{t-1}| x_{t-2}) N^{(t-2)} (x_{t-2}, Y_{t-2} (\cdot)), \label{model-full-02-02-02-01}
\end{align}
where
\begin{align}
E_1^{(t, t-2)} (x_t, y_t (r_t), Y_{t-1} (\cdot)) &= \sum_{\{x_{t-1}'\}} \sum_{\{x_{t-2}'\}} D_1 (x_t, x_{t-1}', y_t (r_t)) D_0 (x_{t-1}', y_{t-1} (r_{t-1}')) \nonumber \\
& \quad \times T_0 (x_{t-1}'| x_{t-2}') p (x_{t-2}'| Y_{t-2} (\cdot)), \\
E_1^{(t-1, t-2)} (x_{t-1}, y_{t-1} (r_{t-1}), Y_{t-2} (\cdot)) &= \sum_{\{x_{t-2}'\}} D_1 (x_{t-1}, x_{t-2}', y_{t-1} (r_{t-1})) p (x_{t-2}'| Y_{t-2} (\cdot)), \\
E_2^{(t, t-1, t-2)} (x_t, y_t (r_t), Y_{t-1} (\cdot)) &= \sum_{\{x_{t-1}'\}} D_1 (x_t, x_{t-1}', y_t (r_t)) E_1^{(t-1, t-2)} (x_{t-1}', y_{t-1} (r_{t-1}'), Y_{t-2} (\cdot)) \nonumber \\
& \quad \times \sum_{\{x_{t-2}'\}} T_0 (x_{t-1}'| x_{t-2}') p (x_{t-2}'| Y_{t-2} (\cdot)).
\end{align}

Equation~\eqref{model-full-02-02-02-01} can be simplified further as
\begin{align}
N^{(t)} (x_t, Y_t (\cdot)) &= \sum_{\{x_{t-1}\}} \sum_{\{x_{t-2}\}} \bigg[ D_0 (x_t, y_t (r_t)) D_0 (x_{t-1}, y_{t-1} (r_{t-1})) \nonumber \\
& \qquad + D_0 (x_t, y_t (r_t)) E_1^{(t-1, t-2)} (x_{t-1}, y_{t-1} (r_{t-1}), Y_{t-2} (\cdot)) \bar{N}^{(t-2)} (Y_{t-2} (\cdot)) \nonumber \\
& \qquad + E_1^{(t, t-2)} (x_t, y_t (r_t), Y_{t-1} (\cdot)) \bar{N}^{(t-2)} (Y_{t-2} (\cdot)) D_0 (x_{t-1}, y_{t-1} (r_{t-1})) \nonumber \\
& \qquad + E_2^{(t, t-1, t-2)} (x_t, y_t (r_t), Y_{t-1} (\cdot)) \Big( \bar{N}^{(t-2)} (Y_{t-2} (\cdot)) \Big)^2 D_0 (x_{t-1}, y_{t-1} (r_{t-1})) \nonumber \\
& \qquad + E_1^{(t, t-2)} (x_t, y_t (r_t), Y_{t-1} (\cdot)) E_1^{(t-1, t-2)} (x_{t-1}, y_{t-1} (r_{t-1}), Y_{t-2} (\cdot)) \Big( \bar{N}^{(t-2)} (Y_{t-2} (\cdot)) \Big)^2 \nonumber \\
& \qquad + E_2^{(t, t-1, t-2)} (x_t, y_t (r_t), Y_{t-1} (\cdot)) E_1^{(t-1, t-2)} (x_{t-1}, y_{t-1} (r_{t-1}), Y_{t-2} (\cdot)) \Big( \bar{N}^{(t-2)} (Y_{t-2} (\cdot)) \Big)^3 \bigg] \nonumber \\
& \quad \times T_0 (x_t| x_{t-1}) T_0 (x_{t-1}| x_{t-2}) N^{(t-2)} (x_{t-2}, Y_{t-2} (\cdot)). \label{model-full-02-02-02}
\end{align}

\subsubsection{First-order perturbative relation between $N^{(t)} (x_t, Y_t (\cdot))$ and $N^{(t-2)} (x_{t-2}, Y_{t-2} (\cdot))$}

We here derive the first-order perturbative relation between $N^{(t)} (x_t, Y_t (\cdot))$ and $N^{(t-2)} (x_{t-2}, Y_{t-2} (\cdot))$.
By neglecting higher order terms in Eq.~\eqref{model-full-02-02-02} with respect to $D_1 (\cdot, \cdot)$, we have
\begin{align}
N^{(t)} (x_t, Y_t (\cdot)) &\approx \sum_{\{x_{t-1}\}} \sum_{\{x_{t-2}\}} \Big[ D_0 (x_t, y_t (r_t)) D_0 (x_{t-1}, y_{t-1} (r_{t-1})) \nonumber \\
& \quad + D_0 (x_t, y_t (r_t)) E_1^{(t-1, t-2)} (x_{t-1}, y_{t-1} (r_{t-1}), Y_{t-2} (\cdot)) \bar{N}^{(t-2)} (Y_{t-2} (\cdot)) \nonumber \\
& \quad + E_1^{(t, t-2)} (x_t, y_t (r_t), Y_{t-1} (\cdot)) \bar{N}^{(t-2)} (Y_{t-2} (\cdot)) D_0 (x_{t-1}, y_{t-1} (r_{t-1})) \Big] \nonumber \\
& \quad \times T_0 (x_t| x_{t-1}) T_0 (x_{t-1}| x_{t-2}) N^{(t-2)} (x_{t-2}, Y_{t-2} (\cdot)) \\
& \approx \sum_{\{x_{t-1}\}} \sum_{\{x_{t-2}\}} \Big( D_0 (x_t, y_t (r_t)) + E_1^{(t, t-2)} (x_t, y_t (r_t), Y_{t-1} (\cdot)) \bar{N}^{(t-2)} (Y_{t-2} (\cdot)) \Big) \nonumber \\
& \quad \times \Big( D_0 (x_{t-1}, y_{t-1} (r_{t-1})) + E_1^{(t-1, t-2)} (x_{t-1}, y_{t-1} (r_{t-1}), Y_{t-2} (\cdot)) \bar{N}^{(t-2)} (Y_{t-2} (\cdot)) \Big) \nonumber \\
& \quad \times T_0 (x_t| x_{t-1}) T_0 (x_{t-1}| x_{t-2}) N^{(t-2)} (x_{t-2}, Y_{t-2} (\cdot)),
\end{align}
where
\begin{align}
E_1^{(t, t-2)} (x_t, y_t (r_t), Y_{t-1} (\cdot)) &= \sum_{\{x_{t-1}'\}} \sum_{\{x_{t-2}'\}} D_1 (x_t, x_{t-1}', y_t (r_t)) D_0 (x_{t-1}', y_{t-1} (r_{t-1}')) \nonumber \\
& \quad \times T_0 (x_{t-1}'| x_{t-2}') p (x_{t-2}'| Y_{t-2} (\cdot)), \\
E_1^{(t-1, t-2)} (x_{t-1}, y_{t-1} (r_{t-1}), Y_{t-2} (\cdot)) &= \sum_{\{x_{t-2}'\}} D_1 (x_{t-1}, x_{t-2}', y_{t-1} (r_{t-1})) p (x_{t-2}'| Y_{t-2} (\cdot)).
\end{align}

\subsubsection{First-order perturbative relation between $N^{(t)} (x_t, Y_t (\cdot))$ and $N^{(t-3)} (x_{t-3}, Y_{t-3} (\cdot))$}

We then derive the first-order perturbative relation between $N^{(t)} (x_t, Y_t (\cdot))$ and $N^{(t-3)} (x_{t-3}, Y_{t-3} (\cdot))$ by using the same procedure.
Then we get
\begin{align}
N^{(t)} (x_t, Y_t (\cdot)) & \approx \sum_{\{x_{t-1}\}} \sum_{\{x_{t-2}\}} \sum_{\{x_{t-3}\}} \nonumber \\
& \quad \times \Big( D_0 (x_t, y_t (r_t)) + E_1^{(t, t-3)} (x_t, y_t (r_t), Y_{t-1} (\cdot)) \bar{N}^{(t-3)} (Y_{t-3} (\cdot)) \Big) \nonumber \\
& \qquad \times \Big( D_0 (x_{t-1}, y_{t-1} (r_{t-1})) + E_1^{(t-1, t-3)} (x_{t-1}, y_{t-1} (r_{t-1}), Y_{t-2} (\cdot)) \bar{N}^{(t-3)} (Y_{t-3} (\cdot)) \Big) \nonumber \\
& \qquad \times \Big( D_0 (x_{t-2}, y_{t-2} (r_{t-2})) + E_1^{(t-2, t-3)} (x_{t-2}, y_{t-2} (r_{t-2}), Y_{t-3} (\cdot)) \bar{N}^{(t-3)} (Y_{t-3} (\cdot)) \Big) \nonumber \\
& \quad \times T_0 (x_t| x_{t-1}) T_0 (x_{t-1}| x_{t-2}) T_0 (x_{t-2}| x_{t-3}) N^{(t-3)} (x_{t-3}, Y_{t-3} (\cdot)),
\end{align}
where
\begin{align}
E_1^{(t, t-3)} (x_t, y_t (r_t), Y_{t-1} (\cdot)) &= \sum_{\{x_{t-1}'\}} \sum_{\{x_{t-2}'\}} \sum_{\{x_{t-3}'\}} D_1 (x_t, x_{t-1}', y_t (r_t)) D_0 (x_{t-1}', y_{t-1} (r_{t-1}')) D_0 (x_{t-2}', y_{t-2} (r_{t-2}')) \nonumber \\
& \quad \times T_0 (x_{t-1}'| x_{t-2}') T_0 (x_{t-2}'| x_{t-3}') p (x_{t-3}'| Y_{t-3} (\cdot)), \\
E_1^{(t-1, t-3)} (x_{t-1}, y_{t-1} (r_{t-1}), Y_{t-2} (\cdot)) &= \sum_{\{x_{t-2}'\}} \sum_{\{x_{t-3}'\}} D_1 (x_{t-1}, x_{t-2}', y_{t-1} (r_{t-1})) D_0 (x_{t-2}', y_{t-2} (r_{t-2}')) \nonumber \\
& \quad \times T_0 (x_{t-2}'| x_{t-3}') p (x_{t-3}'| Y_{t-3} (\cdot)), \\
E_1^{(t-2, t-3)} (x_{t-2}, y_{t-2} (r_{t-2}), Y_{t-3} (\cdot)) &= \sum_{\{x_{t-3}'\}} D_1 (x_{t-2}, x_{t-3}', y_{t-2} (r_{t-2})) p (x_{t-3}'| Y_{t-3} (\cdot)).
\end{align}

\subsubsection{First-order perturbative relation between $N^{(t)} (x_t, Y_t (\cdot))$ and $N^{(0)} (x_0, Y_0 (\cdot))$}

We derive the path integral representation of $N^{(t)} (x_t, Y_t (\cdot))$ in the first-order perturbation by recursively repeating the above procedure.
As a result, we have
\begin{align}
N^{(t)} (x_t, Y_t (\cdot)) &\approx \sum_{\{x_{t-1}\}} \sum_{\{x_{t-2}\}} \dots \sum_{\{x_0\}} \Bigg[ \prod_{i=1}^t \Big( D_0 (x_i, y_i (r_i)) + E_1^{(i, 0)} (x_i, y_i (r_i), Y_{i-1} (\cdot)) \bar{N}^{(0)} \Big) \times T_0 (x_i| x_{i-1}) \Bigg] N^{(0)} (x_0), \label{supp-model-perturbative-02-02}
\end{align}
where
\begin{align}
E_1^{(i, 0)} (x_i, y_i (r_i), Y_{i-1} (\cdot)) &= \sum_{\{x_{i-1}\}} \sum_{\{x_{i-2}\}} \dots \sum_{\{x_0\}} D_1 (x_i, x_{i-1}, y_i (r_i)) \prod_{j=1}^{i-1} \Big[ D_0 (x_j, y_j (r_j)) T_0 (x_j| x_{j-1}) \Big] p (x_0), \label{supp-interaction-term-path-02-01}
\end{align}
$X_t = \{ x_i \}_{i=0}^t$, $Y_t (r) = \{ y_i (r) \}_{i=1}^t$, and $Y_t (\cdot) = \{ y_i (\cdot) \}_{i=1}^t$.
We have also defined
\begin{align}
\bar{N}^{(i)} (Y_i (\cdot)) &= \sum_{\{x_i\}} N^{(i)} (x_i, Y_i (\cdot)), \\
p^{(i)} (x_i| Y_i (\cdot)) &= \frac{N^{(i)} (x_i, Y_i (\cdot))}{\bar{N}^{(i)} (Y_i (\cdot))}.
\end{align}
As mentioned in the main text, we denote $N^{(0)} (x_0, Y_0 (\cdot))$, $p (x_0| Y_0 (\cdot))$, and $\bar{N}^{(0)} (Y_0 (\cdot))$ by $N^{(0)} (x_0)$, $p (x_0)$, and $\bar{N}^{(0)}$, respectively, since $Y_0 (\cdot) = y_0 (\cdot)$, which is the state of the environment at time $t=0$, does not affect the initial state $N^{(0)} (x_0)$.

In the path integral formulation, Eq.~\eqref{supp-model-perturbative-02-02} can be expressed as
\begin{align}
N^{(t)} (x_t, Y_t (\cdot)) &\approx \sum_{\{x_{t-1}\}} \sum_{\{x_{t-2}\}} \dots \sum_{\{x_0\}} \prod_{i=1}^t \Big[ \Big( D_0 (x_i, y_i (r_i)) + E_1^{(i, 0)} (x_i, y_i (r_i), Y_{i-1} (\cdot)) \bar{N}^{(0)} \Big) T_0 (x_i| x_{i-1}) \Big] p (x_0) \bar{N}^{(0)} \\
& = \Braket{ \prod_{i=1}^t \Big( D_0 (x_i, y_i (r_i)) + E_1^{(i, 0)} (x_i, y_i (r_i), Y_{i-1} (\cdot)) \bar{N}^{(0)} \Big) }_{p_\mathrm{f} (X_t)} \bar{N}^{(0)} \\
& = \prod_{i=1}^t \Braket{ \Big( D_0 (x_i, y_i (r_i)) + E_1^{(i, 0)} (x_i, y_i (r_i), Y_{i-1} (\cdot)) \bar{N}^{(0)} \Big) }_{p_\mathrm{f} (X_t)} \bar{N}^{(0)},
\end{align}
where
\begin{align}
p_\mathrm{f} (X_t) &= \Bigg[ \prod_{i=1}^t T_0 (x_i| x_{i-1}) \Bigg] p (x_0),
\end{align}
and $X_t = \{ x_i \}_{i=0}^t$.

We here define the log fitness by
\begin{align}
\Phi_t^\mathrm{tot} (Y_t (\cdot)) &= \ln \frac{\bar{N}^{(t)} (Y_t (\cdot))}{\bar{N}^{(0)}}; \label{supp-log-fitness-02-01-02}
\end{align}
then, it can be rewritten as
\begin{align}
\Phi_t^\mathrm{tot} (Y_t (\cdot)) &= \ln \Braket{ \prod_{i=1}^t \Big( D_0 (x_i, y_i (r_i)) + E_1^{(i, 0)} (x_i, y_i (r_i), Y_{i-1} (\cdot)) \bar{N}^{(0)} \Big) }_{p_\mathrm{f} (X_t)} \\
&= \sum_{i=1}^t \ln \Braket{ \Big( D_0 (x_i, y_i (r_i)) + E_1^{(i, 0)} (x_i, y_i (r_i), Y_{i-1} (\cdot)) \bar{N}^{(0)} \Big) }_{p_\mathrm{f} (X_t)}. \label{supp-log-fitness-02-02-01}
\end{align}
Thus, we have finished the derivation of Eq.~\eqref{first-order-perturbation-02-01}.

%

\subsection{Derivation of Eq.~\eqref{lower-bound-fitness-02-01}} \label{derivation-01-02}

By using Jensen's inequality, we derive Eq.~\eqref{lower-bound-fitness-02-01}.
First, we rewrite Eq.~\eqref{log-fitness-02-01} as follows:
\begin{align}
\Phi_t^\mathrm{tot} (Y_t (\cdot)) &= \ln \frac{\bar{N}^{(t)} (Y_t (\cdot))}{\bar{N}^{(0)}} \\
&= \ln \Braket{ \prod_{i=1}^t \Big( D_0 (x_i, y_i (r_i)) + E_1^{(i, 0)} (x_i, y_i (r_i), Y_{i-1} (\cdot)) \bar{N}^{(0)} \Big) }_{p_\mathrm{f} (X_t)} \\
&= \sum_{i=1}^t \ln \Braket{ \Big( D_0 (x_i, y_i (r_i)) + E_1^{(i, 0)} (x_i, y_i (r_i), Y_{i-1} (\cdot)) \bar{N}^{(0)} \Big) }_{p_\mathrm{f} (X_t)} \\
&= \sum_{i=1}^t \ln \Bigg\langle \Big( D_0 (x_i, y_i (r_i)) + \Braket{ \tilde{E}_1 (X_i, Y_i (\cdot))}_{p_\mathrm{f} (X_{i-1})} \bar{N}^{(0)} \Big) \Bigg\rangle_{p_\mathrm{f} (X_t)} \\
&= \sum_{i=1}^t \ln \Bigg\langle \Big\langle \Big( D_0 (x_i, y_i (r_i)) + \tilde{E}_1 (X_i, Y_i (\cdot)) \bar{N}^{(0)} \Big) \Big\rangle_{p_\mathrm{f} (X_{i-1})} \Bigg\rangle_{p_\mathrm{f} (X_t)} \\
&= \sum_{i=1}^t \ln \Big\langle \Big( D_0 (x_i, y_i (r_i)) + \tilde{E}_1 (X_i, Y_i (\cdot)) \bar{N}^{(0)} \Big) \Big\rangle_{p_\mathrm{f} (X_i)}. \label{inequality-proof-02-01}
\end{align}

Then, we apply Jensen's inequality to Eq.~\eqref{inequality-proof-02-01}, and then we get
\begin{align}
\Phi_t^\mathrm{tot} (Y_t (\cdot)) &\ge \sum_{i=1}^t \Braket{ \ln \Big( D_0 (x_i, y_i (r_i)) + \tilde{E}_1 (X_i, Y_i (\cdot)) \bar{N}^{(0)} \Big) }_{q (X_i)} - \sum_{i=1}^t \mathrm{KL} \Big( q (X_i) \Big\| p_\mathrm{f} (X_i) \Big). \label{inequality-proof-02-02}
\end{align}

We have obtained Eq.~\eqref{lower-bound-fitness-02-01}.

%

\subsection{Derivation of Eq.~\eqref{variational-representation-02-01}} \label{derivation-01-03}

We see that the the equality, Eq.~\eqref{variational-representation-02-01}, is attained by inserting the backward path probability, Eq.~\eqref{backward-path-probability-01}.
By setting
\begin{align}
q (X_i) &= p_\mathrm{b} (X_i| Y_i (\cdot)),
\end{align}
we have
\begin{align}
& \Braket{ \ln \Big( D_0 (x_i) + \tilde{E}_1 (X_i, Y_i (\cdot)) \bar{N}^{(0)} \Big) }_{p_\mathrm{b} (X_i| Y_i (\cdot))} - \mathrm{KL} \Big( p_\mathrm{b} (X_i| Y_i (\cdot)) \Big\| p_\mathrm{f} (X_i) \Big) \nonumber \\
& \quad = \ln \Braket{ \Big( D_0 (x_i) + \tilde{E}_1 (X_i, Y_i (\cdot)) \bar{N}^{(0)} \Big) }_{p_\mathrm{f} (X_i)},
\end{align}
for $i = 1, 2, \dots, t$.
As a result, we get
\begin{align}
& \sum_{i=1}^t \Braket{ \ln \Big( D_0 (x_i) + \tilde{E}_1 (X_i, Y_i (\cdot)) \bar{N}^{(0)} \Big) }_{p_\mathrm{b} (X_i| Y_i (\cdot))} - \sum_{i=1}^t \mathrm{KL} \Big( p_\mathrm{b} (X_i| Y_i (\cdot)) \Big\| p_\mathrm{f} (X_i) \Big) \nonumber \\
& \quad = \sum_{i=1}^t \ln \Braket{ \Big( D_0 (x_i) + \tilde{E}_1 (X_i, Y_i (\cdot)) \bar{N}^{(0)} \Big) }_{p_\mathrm{f} (X_i)}.
\end{align}

Thus, we have Eq.~\eqref{variational-representation-02-01}.

%

\section{Derivations of stochastic thermodynamic relations}

In this appendix, we provide the detailed derivation of the expectation value of the fitness with respect to the environment, Eq.~\eqref{expectation-Phi-02-02}, and its deviations with respect to the forward path probability, Eqs.~\eqref{delta-Phi-01} and \eqref{delta-Phi-02}.

%

\subsection{Derivation of Eq.~\eqref{expectation-Phi-02-02}} \label{derivation-02-01}

The expectation value of the fitness with respect to the environment is easily computed by taking the expectation of Eq.~\eqref{variational-representation-02-01}.
To derive Eq.~\eqref{expectation-Phi-02-02}, we rewrite the second term of the right-hand side of Eq.~\eqref{variational-representation-02-01}.
For $i = 1, 2, \dots, t$, we have
\begin{align}
& \Big\langle \mathrm{KL} \Big( p_\mathrm{b} (X_i| Y_i (\cdot)) \Big\| p_\mathrm{f} (X_i) \Big) \Big\rangle_{p_\mathrm{e} (Y_i (\cdot))} \nonumber \\
& \quad = \sum_{\{X_i\}} \sum_{\{Y_i (\cdot)\}} p_\mathrm{e} (Y_i (\cdot)) p_\mathrm{b} (X_i| Y_i (\cdot)) \ln \frac{p_\mathrm{b} (X_i| Y_i (\cdot))}{p_\mathrm{f} (X_i)} \\
& \quad = \sum_{\{X_i\}} \sum_{\{Y_i (\cdot)\}} p_\mathrm{b} (X_i, Y_i (\cdot)) \ln \frac{p_\mathrm{b} (X_i| Y_i (\cdot))}{p_\mathrm{f} (X_i)} \\
& \quad = \sum_{\{X_i\}} \sum_{\{Y_i (\cdot)\}} p_\mathrm{b} (X_i, Y_i (\cdot)) \ln \frac{p_\mathrm{b} (X_i, Y_i (\cdot))}{p_\mathrm{e} (Y_i (\cdot)) p_\mathrm{f} (X_i)} \\
& \quad = \sum_{\{X_i\}} \sum_{\{Y_i (\cdot)\}} p_\mathrm{b} (X_i, Y_i (\cdot)) \Bigg( \ln \frac{p_\mathrm{b} (X_i, Y_i (\cdot))}{p_\mathrm{e} (Y_i (\cdot)) p_\mathrm{b} (X_i)} + \ln \frac{p_\mathrm{b} (X_i)}{p_\mathrm{f} (X_i)} \Bigg) \\
& \quad = \sum_{\{X_i\}} \sum_{\{Y_i (\cdot)\}} p_\mathrm{b} (X_i, Y_i (\cdot)) \ln \frac{p_\mathrm{b} (X_i, Y_i (\cdot))}{p_\mathrm{e} (Y_i (\cdot)) p_\mathrm{b} (X_i)} + \sum_{\{X_i\}} p_\mathrm{b} (X_i) \ln \frac{p_\mathrm{b} (X_i)}{p_\mathrm{f} (X_i)} \\
& \quad = I_\mathrm{b}^{X_i, Y_i} + \mathrm{KL} \Big( p_\mathrm{b} (X_i) \Big\| p_\mathrm{f} (X_i) \Big).
\end{align}
Thus, we can transform the expectation value of Eq.~\eqref{variational-representation-02-01} with respect to $p_\mathrm{e} (Y_t (\cdot))$ into Eq.~\eqref{expectation-Phi-02-02}.

%

\subsection{Derivations of Eqs.~\eqref{delta-Phi-01} and \eqref{delta-Phi-02}} \label{derivation-02-02}


The deviation of $\delta \hat{\Phi}_t^\mathrm{tot} (Y_t (\cdot))$ defined in Eq.~\eqref{def-dev-fitness-01} with respect to $\hat{p}_\mathrm{b} (X_i| Y_i (\cdot))$ always vanishes since $\hat{\Phi}_t^{\mathrm{tot}} (Y_t (\cdot))$ satisfies Eq.~\eqref{max-variational-representation-02-01}.
Thus we obtain Eq.~\eqref{delta-Phi-01}.
As explained in the main text, by taking the expectation of the left-hand side of Eq.~\eqref{delta-Phi-01} with respect to $p_\mathrm{e}$, we get Eq.~\eqref{delta-Phi-02}.

%

\subsection{Derivation of Eq.~\eqref{maximization-principle-02-10-02}} \label{derivation-02-03}

The proof is given as follows.
\begin{proof}

We first define
\begin{align}
\Lambda (\{ q^{(i, 0)} (X_i| Y_i (\cdot)) \}_{i=1}^t) &= \sum_{i=1}^{t} \Big\langle \ln \Big( D_0 (x_i, y_i (r_i)) + \tilde{E}_1^{(i, 0)} (X_i, Y_i (\cdot)) \bar{N}^{(0)} \Big) \Big\rangle_{q^{(i, 0)} (X_i| Y_i (\cdot)) p_\mathrm{e}^{(i, 0)} (Y_i (\cdot))} - \sum_{i=1}^{t} I^{X_i, Y_i}. \label{proof-02-10-01}
\end{align}
We have, as the derivative of $\Lambda (\{ q^{(i, 0)} (X_i| Y_i (\cdot)) \}_{i=1}^t)$ with respect to $q^{(j, 0)}$,
\begin{align}
& \frac{\delta}{\delta q^{(j, 0)} (X_j| Y_j (\cdot))} \Lambda (\{ q^{(i, 0)} (X_i| Y_i (\cdot)) \}_{i=1}^t) \nonumber \\
& \quad = \Bigg\langle \ln \Big( D_0 (x_j, y_j (r_j)) + \tilde{E}_1^{(j, 0)} (X_j, Y_j (\cdot)) \bar{N}^{(0)} \Big) - \frac{q^{(j, 0)} (X_j| Y_j (\cdot))}{q^{(j, 0)} (X_j)} \Bigg\rangle_{\delta q^{(j, 0)} (X_j| Y_j (\cdot)) p_\mathrm{e}^{(j, 0)} (Y_j (\cdot))}. \label{proof-02-10-02}
\end{align}
The condition that Eq.~\eqref{proof-02-10-02} is zero is expressed as
\begin{align}
q^{(j, 0)} (X_j| Y_j (\cdot)) &\propto \Big( D_0 (x_j, y_j (r_j)) + \tilde{E}_1^{(j, 0)} (X_j, Y_j (\cdot)) \bar{N}^{(0)} \Big) q^{(j, 0)} (X_j).
\end{align}

From Eq.~\eqref{opt-cond-02-02}, we have
\begin{align}
\hat{p}_\mathrm{b} (X_j| Y_j (\cdot)) &= \Big( D_0 (x_j, y_j (r_j)) + \tilde{E}_1^{(j, 0)} (X_j, Y_j (\cdot)) \bar{N}^{(0)} \Big) e^{- \Phi_j (Y_j (\cdot))} \hat{p}_\mathrm{f} (X_j) \\
&\propto \Big( D_0 (x_j, y_j (r_j)) + \tilde{E}_1^{(j, 0)} (X_j, Y_j (\cdot)) \bar{N}^{(0)} \Big) \hat{p}_\mathrm{b} (X_j), \label{proof-02-10-04}
\end{align}
for $j = 1, 2, \dots, t$.

Eq.~\eqref{proof-02-10-04} attains the maximization of Eq.~\eqref{proof-02-10-01}, and thus we have Eq.~\eqref{maximization-principle-02-10-02}.

\end{proof}


%

\section{Derivations of fluctuation relations}

We here elaborate on the derivation of the FRs of interacting population dynamics, Eqs.~\eqref{FR-main-01}, \eqref{fluctuation-relation-main-02-01}, \eqref{fluctuation-relation-main-02-02}, \eqref{fluctuation-relation-kawai-02-01} and \eqref{fluctuation-relation-kawai-02-02}.

%

\subsection{Derivations of Eqs.~\eqref{FR-main-01}, \eqref{fluctuation-relation-main-02-01} and \eqref{fluctuation-relation-main-02-02}} \label{derivation-03-01}

We first define the deviation of the log fitnesses $\hat{\Phi}_j (Y_j (\cdot))$ for $j = 1, 2, \dots, t$ as
\begin{align}
\Delta \hat{\Phi}_j (Y_j (\cdot)) &\coloneqq \hat{\Phi}_j (Y_j (\cdot)) - \Phi_j (Y_j (\cdot));
\end{align}
then, we have the FRs given by
\begin{align}
e^{- \Delta \hat{\Phi}_j (Y_j (\cdot))} &= \frac{\hat{p}_\mathrm{b} (X_j, Y_j (\cdot))}{\hat{p}_\mathrm{f} (X_j)} \frac{p_\mathrm{f} (X_j)}{p_\mathrm{b} (X_j, Y_j (\cdot))}, \label{j-fluctuation-relation-main-02-01} \\
e^{- \Delta \hat{\Phi}_j (Y_j (\cdot))} &= \frac{\hat{p}_\mathrm{b} (Y_j (\cdot)| X_j) p_\mathrm{f} (X_j)}{p_\mathrm{b} (X_j, Y_j (\cdot))}, \label{j-fluctuation-relation-main-02-02} \\
e^{- \Delta \hat{\Phi}_j (Y_j (\cdot))} &= \frac{\Big\langle \hat{p}_\mathrm{b} (Y_j (\cdot)| X_j) \Big\rangle_{p_\mathrm{f} (X_j)}}{p_\mathrm{e} (Y_j (\cdot))}. \label{j-fluctuation-relation-main-02-03}
\end{align}
We prove Eqs.~\eqref{j-fluctuation-relation-main-02-01}, \eqref{j-fluctuation-relation-main-02-02}, and \eqref{j-fluctuation-relation-main-02-03}.
From Eq.~\eqref{backward-path-probability-01}, we have
\begin{align}
\frac{p_\mathrm{b} (X_j| Y_j (\cdot))}{e^{- \Phi_j (Y_j (\cdot))} p_\mathrm{f} (X_j)} &= D_0 (x_j, y_j (r_j)) + \tilde{E}_1 (X_j, Y_j (\cdot)) \bar{N}^{(0)}. \label{proof-02-21-01}
\end{align}
Equation~\eqref{proof-02-21-01} holds for the optimal forward and backward path probabilities; thus, we have the equality given by
\begin{align}
\frac{p_\mathrm{b} (X_j| Y_j (\cdot))}{e^{- \Phi_j (Y_j (\cdot))} p_\mathrm{f} (X_j)} &= \frac{\hat{p}_\mathrm{b} (X_j| Y_j (\cdot))}{e^{- \hat{\Phi}_j (Y_j (\cdot))} \hat{p}_\mathrm{f} (X_j)}.
\end{align}
With simple calculation, we get
\begin{align}
e^{- \Delta \hat{\Phi}_j (Y_j (\cdot))} &= e^{- (\hat{\Phi}_j (Y_j (\cdot)) - \Phi_j (Y_j (\cdot)))} \\
&= \frac{\hat{p}_\mathrm{b} (X_j| Y_j (\cdot))}{\hat{p}_\mathrm{f} (X_j)} \frac{p_\mathrm{f} (X_j)}{p_\mathrm{b} (X_j| Y_j (\cdot))} \label{proof-02-21-04} \\
&= \frac{\hat{p}_\mathrm{b} (X_j, Y_j (\cdot))}{\hat{p}_\mathrm{f} (X_j)} \frac{p_\mathrm{f} (X_j)}{p_\mathrm{b} (X_j, Y_j (\cdot))}. \label{proof-02-21-03}
\end{align}
We have obtained Eq.~\eqref{j-fluctuation-relation-main-02-01}, which is one of the FRs on $X_j$ and $Y_j (\cdot)$.
From Eq.~\eqref{proof-02-21-03}, we obtain
\begin{align}
e^{- \Delta \hat{\Phi}_j (Y_j (\cdot))} &= \frac{\hat{p}_\mathrm{b} (Y_j (\cdot)| X_j) p_\mathrm{f} (X_j)}{p_\mathrm{b} (X_j, Y_j (\cdot))} \frac{\hat{p}_\mathrm{b} (X_j)}{\hat{p}_\mathrm{f} (X_j)}  \label{proof-02-21-05} \\
&= \frac{\hat{p}_\mathrm{b} (Y_j (\cdot)| X_j) p_\mathrm{f} (X_j)}{p_\mathrm{b} (X_j, Y_j (\cdot))}. \label{proof-02-21-07}
\end{align}
Note that from Eq.~\eqref{proof-02-21-05} to Eq.~\eqref{proof-02-21-07}, we have used $\hat{p}_\mathrm{b} (X_j) = \hat{p}_\mathrm{f} (X_j)$ given in Eq.~\eqref{opt-cond-02-02}.
Thus, we have obtained Eq.~\eqref{j-fluctuation-relation-main-02-02}, which is another type of the FRs on $X_j$ and $Y_j (\cdot)$.
Then, we consider the FR on $Y_j (\cdot)$.
From Eq.~\eqref{proof-02-21-07}, we can easily have
\begin{align}
e^{- \Delta \hat{\Phi}_j (Y_j (\cdot))} p_\mathrm{b} (X_j, Y_j (\cdot)) &= \hat{p}_\mathrm{b} (Y_j (\cdot)| X_j) p_\mathrm{f} (X_j). \label{proof-02-21-08}
\end{align}
By summing both sides of Eq.~\eqref{proof-02-21-08} with respect to $X_j$, we have
\begin{align}
e^{- \Delta \hat{\Phi}_j (Y_j (\cdot))} p_\mathrm{e} (Y_j (\cdot)) &= \Big\langle \hat{p}_\mathrm{b} (Y_j (\cdot)| X_j) \Big\rangle_{p_\mathrm{f} (X_j)}.
\end{align}
Thus we have
\begin{align}
e^{- \Delta \hat{\Phi}_j (Y_j (\cdot))} &=  \frac{\Big\langle \hat{p}_\mathrm{b} (Y_j (\cdot)| X_j) \Big\rangle_{p_\mathrm{f} (X_j)}}{p_\mathrm{e} (Y_j (\cdot))}.
\end{align}
We have obtained Eq.~\eqref{j-fluctuation-relation-main-02-03}, which is the FR on $Y_j (\cdot)$.

We finally prove Eqs.~\eqref{FR-main-01}, \eqref{fluctuation-relation-main-02-01} and \eqref{fluctuation-relation-main-02-02} by using Eqs.~\eqref{j-fluctuation-relation-main-02-01}, \eqref{j-fluctuation-relation-main-02-02}, and \eqref{j-fluctuation-relation-main-02-03}, respectively.
By multiplying Eq.~\eqref{j-fluctuation-relation-main-02-01} with respect to $j$, we have
\begin{align}
e^{- \Delta \hat{\Phi}_t^\mathrm{tot} (Y_t (\cdot))} &= \prod_{j=1}^t e^{- \Delta \hat{\Phi}_j (Y_j (\cdot))} \\
&= \prod_{j=1}^t \frac{\hat{p}_\mathrm{b} (X_j| Y_j (\cdot))}{\hat{p}_\mathrm{f} (X_j)} \frac{p_\mathrm{f} (X_j)}{p_\mathrm{b} (X_j| Y_j (\cdot))}.
\end{align}
We have thus obtained Eq.~\eqref{FR-main-01}, which is one of the FRs on $X_t$ and $Y_t (\cdot)$.
Similarly, by multiplying Eq.~\eqref{j-fluctuation-relation-main-02-02} with respect to $j$, we have
\begin{align}
e^{- \Delta \hat{\Phi}_t^\mathrm{tot} (Y_t (\cdot))} &= \prod_{j=1}^t e^{- \Delta \hat{\Phi}_j (Y_j (\cdot))} \\
&= \prod_{j=1}^t \frac{\hat{p}_\mathrm{b} (Y_j (\cdot)| X_j) p_\mathrm{f} (X_j)}{p_\mathrm{b} (X_j, Y_j (\cdot))}.
\end{align}
Thus, we have obtained Eq.~\eqref{fluctuation-relation-main-02-02}, which is another type of the FRs on $X_t$ and $Y_t (\cdot)$.
Again, by multiplying Eq.~\eqref{j-fluctuation-relation-main-02-03} with respect to $j$, we have
\begin{align}
e^{- \Delta \hat{\Phi}_t^\mathrm{tot} (Y_t (\cdot))} &= \prod_{j=1}^t e^{- \Delta \hat{\Phi}_j (Y_j (\cdot))} \\
&= \prod_{j=1}^t \frac{\Big\langle \hat{p}_\mathrm{b} (Y_j (\cdot)| X_j) \Big\rangle_{p_\mathrm{f} (X_j)}}{p_\mathrm{e} (Y_j (\cdot))}.
\end{align}
We have obtained Eq.~\eqref{fluctuation-relation-main-02-01}, which is the FR on $Y_t (\cdot)$.

%

\subsection{Derivations of Eqs.~\eqref{fluctuation-relation-kawai-02-01} and \eqref{fluctuation-relation-kawai-02-02}} \label{derivation-03-03}

We first prove the following Kawai-Parrondo-Broeck type fluctuation relations:
\begin{align}
\Braket{\Delta \hat{\Phi}_j (Y_j (\cdot))}_{p_\mathrm{e} (Y_j (\cdot))} &= \mathrm{KL} \Big( \hat{p}_\mathrm{f} (X_j) \Big\| p_\mathrm{f} (X_j) \Big) - \Big\langle \mathrm{KL} \Big( \hat{p}_\mathrm{b} (X_j| Y_j (\cdot)) \Big\| p_\mathrm{b} (X_j| Y_j (\cdot)) \Big) \Big\rangle_{p_\mathrm{e} (Y_j (\cdot))}, \label{j-fluctuation-relation-kawai-02-01}
\end{align}
and
\begin{align}
\Braket{\Delta \hat{\Phi}_j (Y_j (\cdot))}_{p_\mathrm{e} (Y_j (\cdot))} &= \mathrm{KL} \Big( \hat{p}_\mathrm{f} (X_j) \Big\| p_\mathrm{f} (X_j) \Big) - \mathrm{KL} \Big( \hat{p}_\mathrm{b} (X_j, Y_j (\cdot)) \Big\| p_\mathrm{b} (X_j, Y_j (\cdot)) \Big). \label{j-fluctuation-relation-kawai-02-02}
\end{align}

By taking the logarithm of Eq.~\eqref{proof-02-21-04}, we have
\begin{align}
\Delta \hat{\Phi}_j (Y_j (\cdot)) &= \ln \Bigg( \frac{\hat{p}_\mathrm{f} (X_j)}{\hat{p}_\mathrm{b} (X_j| Y_j (\cdot))} \frac{p_\mathrm{b} (X_j| Y_j (\cdot))}{p_\mathrm{f} (X_j)} \Bigg) \\
&= \ln \Bigg( \frac{\hat{p}_\mathrm{f} (X_j)}{p_\mathrm{f} (X_j)} \frac{p_\mathrm{b} (X_j| Y_j (\cdot))}{\hat{p}_\mathrm{b} (X_j| Y_j (\cdot))} \Bigg) \\
&= \ln \frac{\hat{p}_\mathrm{f} (X_j)}{p_\mathrm{f} (X_j)} - \ln \frac{\hat{p}_\mathrm{b} (X_j| Y_j (\cdot))}{p_\mathrm{b} (X_j| Y_j (\cdot))}. \label{proof-02-22-02}
\end{align}
By taking the expectation of Eq.~\eqref{proof-02-22-02} with respect to $\hat{p}_\mathrm{b} (X_j, Y_j (\cdot))$, we have
\begin{align}
\Braket{\Delta \hat{\Phi}_j (Y_j (\cdot))}_{p_\mathrm{e} (Y_j (\cdot))} &= \Bigg\langle \ln \frac{\hat{p}_\mathrm{f} (X_j)}{p_\mathrm{f} (X_j)} - \ln \frac{\hat{p}_\mathrm{b} (X_j| Y_j (\cdot))}{p_\mathrm{b} (X_j| Y_j (\cdot))} \Bigg\rangle_{\hat{p}_\mathrm{b} (X_j, Y_j (\cdot))} \\
&= \mathrm{KL} \Big( \hat{p}_\mathrm{f} (X_j) \Big\| p_\mathrm{f} (X_j) \Big) - \Big\langle \mathrm{KL} \Big( \hat{p}_\mathrm{b} (X_j| Y_j (\cdot)) \Big\| p_\mathrm{b} (X_j| Y_j (\cdot)) \Big) \Big\rangle_{p_\mathrm{e} (Y_j (\cdot))}.
\end{align}
Thus, we have obtained Eq.~\eqref{j-fluctuation-relation-kawai-02-01}.
With almost the same procedure, we can prove Eq.~\eqref{j-fluctuation-relation-kawai-02-02}.
Summing up Eqs.~\eqref{j-fluctuation-relation-kawai-02-01} and \eqref{j-fluctuation-relation-kawai-02-02} with respect to $j$, respectively, leads to Eqs.~\eqref{fluctuation-relation-kawai-02-01} and \eqref{fluctuation-relation-kawai-02-02}.


%

\end{widetext}

\bibliographystyle{apsrev4-1}
\bibliography{paper-population-dynamics-99-01-bib}

\begin{thebibliography}{27}%
\makeatletter
\providecommand \@ifxundefined [1]{%
 \@ifx{#1\undefined}
}%
\providecommand \@ifnum [1]{%
 \ifnum #1\expandafter \@firstoftwo
 \else \expandafter \@secondoftwo
 \fi
}%
\providecommand \@ifx [1]{%
 \ifx #1\expandafter \@firstoftwo
 \else \expandafter \@secondoftwo
 \fi
}%
\providecommand \natexlab [1]{#1}%
\providecommand \enquote  [1]{``#1''}%
\providecommand \bibnamefont  [1]{#1}%
\providecommand \bibfnamefont [1]{#1}%
\providecommand \citenamefont [1]{#1}%
\providecommand \href@noop [0]{\@secondoftwo}%
\providecommand \href [0]{\begingroup \@sanitize@url \@href}%
\providecommand \@href[1]{\@@startlink{#1}\@@href}%
\providecommand \@@href[1]{\endgroup#1\@@endlink}%
\providecommand \@sanitize@url [0]{\catcode `\\12\catcode `\$12\catcode
  `\&12\catcode `\#12\catcode `\^12\catcode `\_12\catcode `\%12\relax}%
\providecommand \@@startlink[1]{}%
\providecommand \@@endlink[0]{}%
\providecommand \url  [0]{\begingroup\@sanitize@url \@url }%
\providecommand \@url [1]{\endgroup\@href {#1}{\urlprefix }}%
\providecommand \urlprefix  [0]{URL }%
\providecommand \Eprint [0]{\href }%
\providecommand \doibase [0]{http://dx.doi.org/}%
\providecommand \selectlanguage [0]{\@gobble}%
\providecommand \bibinfo  [0]{\@secondoftwo}%
\providecommand \bibfield  [0]{\@secondoftwo}%
\providecommand \translation [1]{[#1]}%
\providecommand \BibitemOpen [0]{}%
\providecommand \bibitemStop [0]{}%
\providecommand \bibitemNoStop [0]{.\EOS\space}%
\providecommand \EOS [0]{\spacefactor3000\relax}%
\providecommand \BibitemShut  [1]{\csname bibitem#1\endcsname}%
\let\auto@bib@innerbib\@empty
\bibitem [{\citenamefont {Mayr}(1999)}]{Mayr01}%
  \BibitemOpen
  \bibfield  {author} {\bibinfo {author} {\bibfnamefont {E.}~\bibnamefont
  {Mayr}},\ }\href@noop {} {\emph {\bibinfo {title} {Systematics and the origin
  of species, from the viewpoint of a zoologist}}}\ (\bibinfo  {publisher}
  {Harvard University Press},\ \bibinfo {year} {1999})\BibitemShut {NoStop}%
\bibitem [{\citenamefont {Gavrilets}(2004)}]{Gavrilets01}%
  \BibitemOpen
  \bibfield  {author} {\bibinfo {author} {\bibfnamefont {S.}~\bibnamefont
  {Gavrilets}},\ }\href@noop {} {\emph {\bibinfo {title} {Fitness landscapes
  and the origin of species (MPB-41)}}},\ Vol.~\bibinfo {volume} {41}\
  (\bibinfo  {publisher} {Princeton University Press},\ \bibinfo {year}
  {2004})\BibitemShut {NoStop}%
\bibitem [{\citenamefont {Mesz\'ena}\ \emph {et~al.}(2005)\citenamefont
  {Mesz\'ena}, \citenamefont {Gyllenberg}, \citenamefont {Jacobs},\ and\
  \citenamefont {Metz}}]{Meszena01}%
  \BibitemOpen
  \bibfield  {author} {\bibinfo {author} {\bibfnamefont {G.}~\bibnamefont
  {Mesz\'ena}}, \bibinfo {author} {\bibfnamefont {M.}~\bibnamefont
  {Gyllenberg}}, \bibinfo {author} {\bibfnamefont {F.~J.}\ \bibnamefont
  {Jacobs}}, \ and\ \bibinfo {author} {\bibfnamefont {J.~A.~J.}\ \bibnamefont
  {Metz}},\ }\href {\doibase 10.1103/PhysRevLett.95.078105} {\bibfield
  {journal} {\bibinfo  {journal} {Phys. Rev. Lett.}\ }\textbf {\bibinfo
  {volume} {95}},\ \bibinfo {pages} {078105} (\bibinfo {year}
  {2005})}\BibitemShut {NoStop}%
\bibitem [{\citenamefont {Thieme}(2018)}]{Thieme01}%
  \BibitemOpen
  \bibfield  {author} {\bibinfo {author} {\bibfnamefont {H.~R.}\ \bibnamefont
  {Thieme}},\ }\href@noop {} {\emph {\bibinfo {title} {Mathematics in
  population biology}}}\ (\bibinfo  {publisher} {Princeton University Press},\
  \bibinfo {year} {2018})\BibitemShut {NoStop}%
\bibitem [{\citenamefont {Hofbauer}\ and\ \citenamefont
  {Sigmund}(1998)}]{Hofbauer01}%
  \BibitemOpen
  \bibfield  {author} {\bibinfo {author} {\bibfnamefont {J.}~\bibnamefont
  {Hofbauer}}\ and\ \bibinfo {author} {\bibfnamefont {K.}~\bibnamefont
  {Sigmund}},\ }\href@noop {} {\emph {\bibinfo {title} {Evolutionary games and
  population dynamics}}}\ (\bibinfo  {publisher} {Cambridge university press},\
  \bibinfo {year} {1998})\BibitemShut {NoStop}%
\bibitem [{\citenamefont {Turchin}(2003)}]{Turchin01}%
  \BibitemOpen
  \bibfield  {author} {\bibinfo {author} {\bibfnamefont {P.}~\bibnamefont
  {Turchin}},\ }\href@noop {} {\emph {\bibinfo {title} {Complex population
  dynamics: a theoretical/empirical synthesis}}},\ Vol.~\bibinfo {volume} {35}\
  (\bibinfo  {publisher} {Princeton university press},\ \bibinfo {year}
  {2003})\BibitemShut {NoStop}%
\bibitem [{\citenamefont {Hartl}\ \emph {et~al.}(1997)\citenamefont {Hartl},
  \citenamefont {Clark},\ and\ \citenamefont {Clark}}]{Hartl01}%
  \BibitemOpen
  \bibfield  {author} {\bibinfo {author} {\bibfnamefont {D.~L.}\ \bibnamefont
  {Hartl}}, \bibinfo {author} {\bibfnamefont {A.~G.}\ \bibnamefont {Clark}}, \
  and\ \bibinfo {author} {\bibfnamefont {A.~G.}\ \bibnamefont {Clark}},\
  }\href@noop {} {\emph {\bibinfo {title} {Principles of population
  genetics}}},\ Vol.\ \bibinfo {volume} {116}\ (\bibinfo  {publisher} {Sinauer
  associates Sunderland},\ \bibinfo {year} {1997})\BibitemShut {NoStop}%
\bibitem [{\citenamefont {Lande}\ \emph {et~al.}(2003)\citenamefont {Lande},
  \citenamefont {Engen},\ and\ \citenamefont {Saether}}]{Lande01}%
  \BibitemOpen
  \bibfield  {author} {\bibinfo {author} {\bibfnamefont {R.}~\bibnamefont
  {Lande}}, \bibinfo {author} {\bibfnamefont {S.}~\bibnamefont {Engen}}, \ and\
  \bibinfo {author} {\bibfnamefont {B.-E.}\ \bibnamefont {Saether}},\
  }\href@noop {} {\emph {\bibinfo {title} {Stochastic population dynamics in
  ecology and conservation}}}\ (\bibinfo  {publisher} {Oxford University Press
  on Demand},\ \bibinfo {year} {2003})\BibitemShut {NoStop}%
\bibitem [{\citenamefont {Donaldson-Matasci}\ \emph {et~al.}(2010)\citenamefont
  {Donaldson-Matasci}, \citenamefont {Bergstrom},\ and\ \citenamefont
  {Lachmann}}]{Donaldson01}%
  \BibitemOpen
  \bibfield  {author} {\bibinfo {author} {\bibfnamefont {M.~C.}\ \bibnamefont
  {Donaldson-Matasci}}, \bibinfo {author} {\bibfnamefont {C.~T.}\ \bibnamefont
  {Bergstrom}}, \ and\ \bibinfo {author} {\bibfnamefont {M.}~\bibnamefont
  {Lachmann}},\ }\href@noop {} {\bibfield  {journal} {\bibinfo  {journal}
  {Oikos}\ }\textbf {\bibinfo {volume} {119}},\ \bibinfo {pages} {219}
  (\bibinfo {year} {2010})}\BibitemShut {NoStop}%
\bibitem [{\citenamefont {Rivoire}\ and\ \citenamefont
  {Leibler}(2011)}]{Rivoire01}%
  \BibitemOpen
  \bibfield  {author} {\bibinfo {author} {\bibfnamefont {O.}~\bibnamefont
  {Rivoire}}\ and\ \bibinfo {author} {\bibfnamefont {S.}~\bibnamefont
  {Leibler}},\ }\href@noop {} {\bibfield  {journal} {\bibinfo  {journal}
  {Journal of Statistical Physics}\ }\textbf {\bibinfo {volume} {142}},\
  \bibinfo {pages} {1124} (\bibinfo {year} {2011})}\BibitemShut {NoStop}%
\bibitem [{\citenamefont {Rivoire}(2016)}]{Rivoire02}%
  \BibitemOpen
  \bibfield  {author} {\bibinfo {author} {\bibfnamefont {O.}~\bibnamefont
  {Rivoire}},\ }\href {\doibase 10.1007/s10955-015-1381-z} {\bibfield
  {journal} {\bibinfo  {journal} {Journal of Statistical Physics}\ }\textbf
  {\bibinfo {volume} {162}},\ \bibinfo {pages} {1324} (\bibinfo {year}
  {2016})}\BibitemShut {NoStop}%
\bibitem [{\citenamefont {Jarzynski}(1997{\natexlab{a}})}]{Jarzynski01}%
  \BibitemOpen
  \bibfield  {author} {\bibinfo {author} {\bibfnamefont {C.}~\bibnamefont
  {Jarzynski}},\ }\href {\doibase 10.1103/PhysRevLett.78.2690} {\bibfield
  {journal} {\bibinfo  {journal} {Phys. Rev. Lett.}\ }\textbf {\bibinfo
  {volume} {78}},\ \bibinfo {pages} {2690} (\bibinfo {year}
  {1997}{\natexlab{a}})}\BibitemShut {NoStop}%
\bibitem [{\citenamefont {Crooks}(1999)}]{Crooks01}%
  \BibitemOpen
  \bibfield  {author} {\bibinfo {author} {\bibfnamefont {G.~E.}\ \bibnamefont
  {Crooks}},\ }\href {\doibase 10.1103/PhysRevE.60.2721} {\bibfield  {journal}
  {\bibinfo  {journal} {Phys. Rev. E}\ }\textbf {\bibinfo {volume} {60}},\
  \bibinfo {pages} {2721} (\bibinfo {year} {1999})}\BibitemShut {NoStop}%
\bibitem [{\citenamefont {Seifert}(2012)}]{Seifert01}%
  \BibitemOpen
  \bibfield  {author} {\bibinfo {author} {\bibfnamefont {U.}~\bibnamefont
  {Seifert}},\ }\href {http://stacks.iop.org/0034-4885/75/i=12/a=126001}
  {\bibfield  {journal} {\bibinfo  {journal} {Reports on Progress in Physics}\
  }\textbf {\bibinfo {volume} {75}},\ \bibinfo {pages} {126001} (\bibinfo
  {year} {2012})}\BibitemShut {NoStop}%
\bibitem [{\citenamefont {Kobayashi}\ and\ \citenamefont
  {Sughiyama}(2015)}]{Kobayashi03}%
  \BibitemOpen
  \bibfield  {author} {\bibinfo {author} {\bibfnamefont {T.~J.}\ \bibnamefont
  {Kobayashi}}\ and\ \bibinfo {author} {\bibfnamefont {Y.}~\bibnamefont
  {Sughiyama}},\ }\href@noop {} {\bibfield  {journal} {\bibinfo  {journal}
  {Physical review letters}\ }\textbf {\bibinfo {volume} {115}},\ \bibinfo
  {pages} {238102} (\bibinfo {year} {2015})}\BibitemShut {NoStop}%
\bibitem [{\citenamefont {Kobayashi}\ and\ \citenamefont
  {Sughiyama}(2017)}]{Kobayashi04}%
  \BibitemOpen
  \bibfield  {author} {\bibinfo {author} {\bibfnamefont {T.~J.}\ \bibnamefont
  {Kobayashi}}\ and\ \bibinfo {author} {\bibfnamefont {Y.}~\bibnamefont
  {Sughiyama}},\ }\href@noop {} {\bibfield  {journal} {\bibinfo  {journal}
  {Physical Review E}\ }\textbf {\bibinfo {volume} {96}},\ \bibinfo {pages}
  {012402} (\bibinfo {year} {2017})}\BibitemShut {NoStop}%
\bibitem [{\citenamefont {Tsoularis}\ and\ \citenamefont
  {Wallace}(2002)}]{Tsoularis01}%
  \BibitemOpen
  \bibfield  {author} {\bibinfo {author} {\bibfnamefont {A.}~\bibnamefont
  {Tsoularis}}\ and\ \bibinfo {author} {\bibfnamefont {J.}~\bibnamefont
  {Wallace}},\ }\href@noop {} {\bibfield  {journal} {\bibinfo  {journal}
  {Mathematical biosciences}\ }\textbf {\bibinfo {volume} {179}},\ \bibinfo
  {pages} {21} (\bibinfo {year} {2002})}\BibitemShut {NoStop}%
\bibitem [{\citenamefont {Abrikosov}\ \emph {et~al.}(2012)\citenamefont
  {Abrikosov}, \citenamefont {Gorkov},\ and\ \citenamefont
  {Dzyaloshinski}}]{Abrikosov01}%
  \BibitemOpen
  \bibfield  {author} {\bibinfo {author} {\bibfnamefont {A.~A.}\ \bibnamefont
  {Abrikosov}}, \bibinfo {author} {\bibfnamefont {L.~P.}\ \bibnamefont
  {Gorkov}}, \ and\ \bibinfo {author} {\bibfnamefont {I.~E.}\ \bibnamefont
  {Dzyaloshinski}},\ }\href@noop {} {\  (\bibinfo {year} {2012})}\BibitemShut
  {NoStop}%
\bibitem [{\citenamefont {Fetter}\ and\ \citenamefont
  {Walecka}(2003)}]{Fetter01}%
  \BibitemOpen
  \bibfield  {author} {\bibinfo {author} {\bibfnamefont {A.~L.}\ \bibnamefont
  {Fetter}}\ and\ \bibinfo {author} {\bibfnamefont {J.~D.}\ \bibnamefont
  {Walecka}},\ }\href@noop {} {\emph {\bibinfo {title} {Quantum theory of
  many-particle systems}}}\ (\bibinfo  {publisher} {Courier Corporation},\
  \bibinfo {year} {2003})\BibitemShut {NoStop}%
\bibitem [{\citenamefont {Kawai}\ \emph {et~al.}(2007)\citenamefont {Kawai},
  \citenamefont {Parrondo},\ and\ \citenamefont {Van~den Broeck}}]{Kawai01}%
  \BibitemOpen
  \bibfield  {author} {\bibinfo {author} {\bibfnamefont {R.}~\bibnamefont
  {Kawai}}, \bibinfo {author} {\bibfnamefont {J.}~\bibnamefont {Parrondo}}, \
  and\ \bibinfo {author} {\bibfnamefont {C.}~\bibnamefont {Van~den Broeck}},\
  }\href@noop {} {\bibfield  {journal} {\bibinfo  {journal} {Physical review
  letters}\ }\textbf {\bibinfo {volume} {98}},\ \bibinfo {pages} {080602}
  (\bibinfo {year} {2007})}\BibitemShut {NoStop}%
\bibitem [{\citenamefont {Jarzynski}(2011)}]{Jarzynski02}%
  \BibitemOpen
  \bibfield  {author} {\bibinfo {author} {\bibfnamefont {C.}~\bibnamefont
  {Jarzynski}},\ }\href {\doibase 10.1146/annurev-conmatphys-062910-140506}
  {\bibfield  {journal} {\bibinfo  {journal} {Annual Review of Condensed Matter
  Physics}\ }\textbf {\bibinfo {volume} {2}},\ \bibinfo {pages} {329} (\bibinfo
  {year} {2011})},\ \Eprint
  {http://arxiv.org/abs/https://doi.org/10.1146/annurev-conmatphys-062910-140506}
  {https://doi.org/10.1146/annurev-conmatphys-062910-140506} \BibitemShut
  {NoStop}%
\bibitem [{\citenamefont {Jarzynski}(1997{\natexlab{b}})}]{Jarzynski03}%
  \BibitemOpen
  \bibfield  {author} {\bibinfo {author} {\bibfnamefont {C.}~\bibnamefont
  {Jarzynski}},\ }\href {\doibase 10.1103/PhysRevE.56.5018} {\bibfield
  {journal} {\bibinfo  {journal} {Phys. Rev. E}\ }\textbf {\bibinfo {volume}
  {56}},\ \bibinfo {pages} {5018} (\bibinfo {year}
  {1997}{\natexlab{b}})}\BibitemShut {NoStop}%
\bibitem [{\citenamefont {Seifert}(2017)}]{Seifert02}%
  \BibitemOpen
  \bibfield  {author} {\bibinfo {author} {\bibfnamefont {U.}~\bibnamefont
  {Seifert}},\ }\href@noop {} {\bibfield  {journal} {\bibinfo  {journal}
  {Physica A: Statistical Mechanics and its Applications}\ } (\bibinfo {year}
  {2017})}\BibitemShut {NoStop}%
\bibitem [{\citenamefont {Seifert}(2005)}]{Seifert03}%
  \BibitemOpen
  \bibfield  {author} {\bibinfo {author} {\bibfnamefont {U.}~\bibnamefont
  {Seifert}},\ }\href {\doibase 10.1103/PhysRevLett.95.040602} {\bibfield
  {journal} {\bibinfo  {journal} {Phys. Rev. Lett.}\ }\textbf {\bibinfo
  {volume} {95}},\ \bibinfo {pages} {040602} (\bibinfo {year}
  {2005})}\BibitemShut {NoStop}%
\bibitem [{\citenamefont {Kurchan}(1998)}]{Kurchan01}%
  \BibitemOpen
  \bibfield  {author} {\bibinfo {author} {\bibfnamefont {J.}~\bibnamefont
  {Kurchan}},\ }\href {http://stacks.iop.org/0305-4470/31/i=16/a=003}
  {\bibfield  {journal} {\bibinfo  {journal} {Journal of Physics A:
  Mathematical and General}\ }\textbf {\bibinfo {volume} {31}},\ \bibinfo
  {pages} {3719} (\bibinfo {year} {1998})}\BibitemShut {NoStop}%
\bibitem [{\citenamefont {Miyahara}\ and\ \citenamefont
  {Aihara}(2018)}]{Miyahara07}%
  \BibitemOpen
  \bibfield  {author} {\bibinfo {author} {\bibfnamefont {H.}~\bibnamefont
  {Miyahara}}\ and\ \bibinfo {author} {\bibfnamefont {K.}~\bibnamefont
  {Aihara}},\ }\href {\doibase 10.1103/PhysRevE.98.042138} {\bibfield
  {journal} {\bibinfo  {journal} {Phys. Rev. E}\ }\textbf {\bibinfo {volume}
  {98}},\ \bibinfo {pages} {042138} (\bibinfo {year} {2018})}\BibitemShut
  {NoStop}%
\bibitem [{\citenamefont {Kermack}\ and\ \citenamefont
  {McKendrick}(1927)}]{Kermack01}%
  \BibitemOpen
  \bibfield  {author} {\bibinfo {author} {\bibfnamefont {W.~O.}\ \bibnamefont
  {Kermack}}\ and\ \bibinfo {author} {\bibfnamefont {A.~G.}\ \bibnamefont
  {McKendrick}},\ }\href@noop {} {\bibfield  {journal} {\bibinfo  {journal}
  {Proceedings of the royal society of london. Series A, Containing papers of a
  mathematical and physical character}\ }\textbf {\bibinfo {volume} {115}},\
  \bibinfo {pages} {700} (\bibinfo {year} {1927})}\BibitemShut {NoStop}%
\end{thebibliography}%

\end{document}